\documentclass[journal]{IEEEtran}
\usepackage[utf8]{inputenc}
\usepackage{etex}
\usepackage{amsmath, amssymb, amsfonts, bm, nccmath}                        % better math environment
\usepackage{booktabs}                       % better tables
\usepackage{graphicx}
\usepackage{amssymb, stfloats, amssymb, amsfonts, rotating, acronym}
\usepackage{multirow}
\usepackage{relsize}
\usepackage{array}
\usepackage{tabularx}
\usepackage{tikz}
\usetikzlibrary{patterns}
\usetikzlibrary{arrows,positioning,shapes.geometric,spy,matrix}
\usetikzlibrary{backgrounds}
\usepackage{filecontents}
\usepackage{mathtools}
\usepackage{color}
\usepackage{xcolor}
\usepackage{pgfplots}
\pgfplotsset{compat=newest}
\usepgfplotslibrary{groupplots}
\usepackage{cite}
\usepackage{adjustbox}
\usepackage{colortbl}
\usepackage{lipsum}
\usepackage[linesnumbered,algoruled,boxed,vlined]{algorithm2e}
\usepackage[normalem]{ulem}
\DeclareMathOperator*{\sgn}{sgn}
\DeclareMathOperator*{\erfc}{erfc}
\DeclarePairedDelimiter\abs{\lvert}{\rvert}

\newcommand{\fixme}[2]{\ifx&#2&{\leavevmode\color{red}#1}\else{\leavevmode\color{red}FIXME\{}#1{\leavevmode\color{red}\}}\footnote{{\leavevmode\color{red}#2}}\PackageWarning{Fixme}{#1: #2}\fi}

\newcommand{\pluseq}{\mathrel{+}=}

\DeclarePairedDelimiterX\Basics[1](){ #1}

\definecolor{bblue}{HTML}{4F81BD}
\definecolor{rred}{HTML}{C0504D}
\definecolor{ggreen}{HTML}{9BBB59}
\definecolor{ppurple}{HTML}{9F4C7C}

\definecolor{ocre}{RGB}{0, 157, 224} % Define the orange color used for highlighting throughout the book
\definecolor{Paired-2}{RGB}{166,206,227}
\definecolor{Paired-1}{RGB}{31,120,180}
\definecolor{Paired-4}{RGB}{178,223,138}
\definecolor{Paired-3}{RGB}{51,160,44}
\definecolor{Paired-6}{RGB}{251,154,153}
\definecolor{Paired-5}{RGB}{227,26,28}
\definecolor{Paired-8}{RGB}{253,191,111}
\definecolor{Paired-7}{RGB}{255,127,0}
\definecolor{Paired-10}{RGB}{202,178,214}
\definecolor{Paired-9}{RGB}{106,61,154}
\definecolor{Paired-12}{RGB}{255,255,153}
\definecolor{Paired-11}{RGB}{177,89,40}
\definecolor{Accent-1}{RGB}{127,201,127}
\definecolor{Accent-2}{RGB}{190,174,212}
\definecolor{Accent-3}{RGB}{253,192,134}
\definecolor{Accent-4}{RGB}{255,255,153}
\definecolor{Accent-5}{RGB}{56,108,176}
\definecolor{Accent-6}{RGB}{240,2,127}
\definecolor{Accent-7}{RGB}{191,91,23}
\definecolor{Accent-8}{RGB}{102,102,102}
\definecolor{Spectral-1}{RGB}{158,1,66}
\definecolor{Spectral-2}{RGB}{213,62,79}
\definecolor{Spectral-3}{RGB}{244,109,67}
\definecolor{Spectral-4}{RGB}{253,174,97}
\definecolor{Spectral-5}{RGB}{254,224,139}
\definecolor{Spectral-6}{RGB}{255,255,191}
\definecolor{Spectral-7}{RGB}{230,245,152}
\definecolor{Spectral-8}{RGB}{171,221,164}
\definecolor{Spectral-9}{RGB}{102,194,165}
\definecolor{Spectral-10}{RGB}{50,136,189}
\definecolor{Spectral-11}{RGB}{94,79,162}
\definecolor{Set1-1}{RGB}{228,26,28}
\definecolor{Set1-2}{RGB}{55,126,184}
\definecolor{Set1-3}{RGB}{77,175,74}
\definecolor{Set1-4}{RGB}{152,78,163}
\definecolor{Set1-5}{RGB}{255,127,0}
\definecolor{Set1-6}{RGB}{255,255,51}
\definecolor{Set1-7}{RGB}{166,86,40}
\definecolor{Set1-8}{RGB}{247,129,191}
\definecolor{Set1-9}{RGB}{153,153,153}
\definecolor{Set2-1}{RGB}{102,194,165}
\definecolor{Set2-2}{RGB}{252,141,98}
\definecolor{Set2-3}{RGB}{141,160,203}
\definecolor{Set2-4}{RGB}{231,138,195}
\definecolor{Set2-5}{RGB}{166,216,84}
\definecolor{Set2-6}{RGB}{255,217,47}
\definecolor{Set2-7}{RGB}{229,196,148}
\definecolor{Set2-8}{RGB}{179,179,179}
\definecolor{Dark2-1}{RGB}{27,158,119}
\definecolor{Dark2-2}{RGB}{217,95,2}
\definecolor{Dark2-3}{RGB}{117,112,179}
\definecolor{Dark2-4}{RGB}{231,41,138}
\definecolor{Dark2-5}{RGB}{102,166,30}
\definecolor{Dark2-6}{RGB}{230,171,2}
\definecolor{Dark2-7}{RGB}{166,118,29}
\definecolor{Dark2-8}{RGB}{102,102,102}
\definecolor{Reds-1}{RGB}{255,245,240}
\definecolor{Reds-2}{RGB}{254,224,210}
\definecolor{Reds-3}{RGB}{252,187,161}
\definecolor{Reds-4}{RGB}{252,146,114}
\definecolor{Reds-5}{RGB}{251,106,74}
\definecolor{Reds-6}{RGB}{239,59,44}
\definecolor{Reds-7}{RGB}{203,24,29}
\definecolor{Reds-8}{RGB}{165,15,21}
\definecolor{Reds-9}{RGB}{103,0,13}
\definecolor{Greens-1}{RGB}{247,252,245}
\definecolor{Greens-2}{RGB}{229,245,224}
\definecolor{Greens-3}{RGB}{199,233,192}
\definecolor{Greens-4}{RGB}{161,217,155}
\definecolor{Greens-5}{RGB}{116,196,118}
\definecolor{Greens-6}{RGB}{65,171,93}
\definecolor{Greens-7}{RGB}{35,139,69}
\definecolor{Greens-8}{RGB}{0,109,44}
\definecolor{Greens-9}{RGB}{0,68,27}
\definecolor{Blues-1}{RGB}{247,251,255}
\definecolor{Blues-2}{RGB}{222,235,247}
\definecolor{Blues-3}{RGB}{198,219,239}
\definecolor{Blues-4}{RGB}{158,202,225}
\definecolor{Blues-5}{RGB}{107,174,214}
\definecolor{Blues-6}{RGB}{66,146,198}
\definecolor{Blues-7}{RGB}{33,113,181}
\definecolor{Blues-8}{RGB}{8,81,156}
\definecolor{Blues-9}{RGB}{8,48,107}
\definecolor{awesome-skyblue}{HTML}{0395DE}
\definecolor{bg}{HTML}{282828}
\definecolor{rosso}{RGB}{220,57,18}
\definecolor{giallo}{RGB}{255,153,0}
\definecolor{blu}{RGB}{102,140,217}
\definecolor{verde}{RGB}{16,150,24}
\definecolor{viola}{RGB}{153,0,153}

\definecolor{bleuUni}{RGB}{0, 157, 224}
\definecolor{marronUni}{RGB}{68, 58, 49}

\begin{document}

\bstctlcite{IEEEexample:BSTcontrol}

\title{Practical Dynamic SC-Flip Polar Decoders: \\ Algorithm and Implementation}

\pgfmathdeclarefunction{gauss}{2}{%
  \pgfmathparse{1/(#2*sqrt(2*pi))*exp(-((x-#1)^2)/(2*#2^2))}%
}

\author{Furkan~Ercan,~\IEEEmembership{Student~Member,~IEEE,}
        Thibaud~Tonnellier,
        Nghia~Doan,~\IEEEmembership{Student~Member,~IEEE,}
        and~Warren~J.~Gross,~\IEEEmembership{Senior Member,~IEEE}% <-this % stops a space
\thanks{F.~Ercan, T.~Tonnellier, N.~Doan and W.~J.~Gross are with the Department of Electrical and Computer Engineering, McGill University, Montr\'eal, Qu\'ebec, Canada. (e-mail: furkan.ercan@mail.mcgill.ca, thibaud.tonnellier@mcgill.ca, nghia.doan@mail.mcgill.ca, warren.gross@mcgill.ca.)}% <-this % stops a space
\thanks{Part of this work has been published in ICASSP 2020.}
}

\markboth{Journal of \LaTeX\ Class Files,~Vol.~14, No.~8, August~2015}%
{Shell \MakeLowercase{\textit{et al.}}: Bare Demo of IEEEtran.cls for IEEE Journals}


\maketitle

\begin{abstract}%IEEE TSP: 150-250 words
SC-Flip (SCF) is a low-complexity polar code decoding algorithm with improved performance, and is an alternative to high-complexity (CRC)-aided SC-List (CA-SCL) decoding. However, the performance improvement of SCF is limited since it can correct up to only one channel error ($\omega=1$). Dynamic SCF (DSCF) algorithm tackles this problem by tackling multiple errors ($\omega \geq 1$), but it requires logarithmic and exponential computations, which make it infeasible for practical applications. In this work, we propose simplifications and approximations to make DSCF practically feasible. First, we reduce the transcendental computations of DSCF decoding to a constant approximation. Then, we show how to incorporate special node decoding techniques into DSCF algorithm, creating the Fast-DSCF decoding. Next, we reduce the search span within the special nodes to further reduce the computational complexity. Following, we describe a hardware architecture for the Fast-DSCF decoder, in which we introduce additional simplifications such as metric normalization and sorter length reduction. All the simplifications and approximations are shown to have minimal impact on the error-correction performance, and the reported Fast-DSCF decoder is the only SCF-based architecture that can correct multiple errors. The Fast-DSCF decoders synthesized using TSMC $65$nm CMOS technology can achieve a $1.25$, $1.06$ and $0.93$ Gbps throughput for $\omega \in \{1,2,3\}$, respectively. Compared to the state-of-the-art fast CA-SCL decoders with equivalent FER performance, the proposed decoders are up to $5.8\times$ more area-efficient. Finally, observations at energy dissipation indicate that the Fast-DSCF is more energy-efficient than its CA-SCL-based counterparts.

\end{abstract}

\begin{IEEEkeywords}
Polar codes, 5G, energy efficiency, Dynamic SCFlip, wireless communications, hardware implementation.
\end{IEEEkeywords}

\IEEEpeerreviewmaketitle

\vspace{-1em}

\section{Introduction}\label{sec:intro}

The $5^{\text{th}}$ generation wireless mobile communications standard (5G) creates a vast infrastructure that enhances the existing communications platforms and enables new technologies. Among the 5G use cases, massive machine-type communications (mMTC) \cite{5g_mmtc_IEEEComm2016} prioritize enhanced connectivity and energy efficiency. 

Polar codes are a class of forward error-correcting codes that asymptotically achieve the channel capacity \cite{arikan09}. They have been selected as the coding scheme for the control channel for 5G eMBB \cite{38.212}, and are being evaluated for 5G URLLC and mMTC use cases \cite{sharma2017polar,sybis2016channel}. Even though the successive cancellation (SC) decoding algorithm of polar codes enables to prove the capacity achieving property, its error-correction performance is mediocre at practical code lengths. 

In order to improve the error correction performance of polar codes, SC-List (SCL) decoding was proposed \cite{TalList}. SCL uses $L$ SC decoders in parallel to maintain a list of candidate codewords which improves the error-correction performance at the cost of increased implementation complexity \cite{ercan-allerton}. SC-Flip (SCF) decoding \cite{SCFlip14} is another SC-based polar decoder algorithm that uses several SC decoding attempts when an initial SC decoding fails due to a single channel-induced error. Compared to SC decoding, SCF has improved error-correction performance at the cost of variable decoding latency. The average computational complexity of SCF decoding is similar to that of SC decoding at medium-to-high signal-to-noise ratio (SNR) regions. However, the improved performance with the SCF decoding is limited and can only match to its SCL counterparts with small list sizes. The limited performance improvement of the SCF is due to two main issues. The first problem is that SCF cannot correct more than one channel-induced error. The second problem is that the metric used to identify the error is suboptimal. 

Dynamic SC-Flip (DSCF) decoding \cite{SCFlip17-jour} proposes a solution to address both of these problems, by extending the search to more than one channel-induced errors, and by proposing an enhanced metric that is significantly more efficient on locating the erroneous locations in the codeword. In return, the logarithmic and exponential calculations involved in the DSCF decoding make it challenging for practical hardware implementations.

Our goal in this work is to make the DSCF algorithm practically feasible, so that it can be implemented in hardware at low cost to become an alternative for existing high-performance polar decoder architectures. State-of-the-art decoder architectures for polar codes either require substantial amount of resources (e.g. Fast-SSCL decoding \cite{fastSSCL-TSP}), or have limited error-correction performance improvement (e.g. Fast-SCF decoding \cite{FastSCF-TCAS-I}). Accordingly, our contributions are summarized as follows:
\begin{itemize}
\item First, we show that the logarithmic and exponential computations in the DSCF algorithm can be replaced by a simple constant approximation. We show that the proposed approximation does not incur any significant loss in error-correction performance.
\item Then, we propose novel methods to implement decoding of special nodes under DSCF algorithm. We reformulate the original computations of the DSCF decoding to accomodate special nodes, and we show that it is possible to maintain similar error-correction performance. Moreover, we show that the achievable error-correction performance can in fact be improved with one of the special nodes. 
\item We show how to reduce the computational complexity further associated with two special nodes. Using a mathematical framework, we first find the theoretical frame-error rate (FER) for these special nodes with and without error-correction. Then, we show the achievable performance approximations when the computational effort in these nodes are intentionally reduced. Under the light of these findings, we limit the computational effort in the hardware architecture that follows. 
\item Finally, we show how to implement the proposed Fast-DSCF algorithm in hardware. The proposed hardware takes advantage of all the proposed simplifications. In addition, we present further simplifications, such as metric normalization and sorter length reduction. There are several SCF-based algorithms that describe multiple bit-flipping operations, but to the best of our knowledge none of them describe a hardware architecture. Therefore, the proposed Fast-DSCF decoder is the first reported SCF-based decoder architecture that can correct more than a single channel-induced error. 
\end{itemize}

Simulation results show that Fast-DSCF decoder using all the simplifications, approximations and quantizations maintains similar error-correction performance to the baseline DSCF algorithm. Moreover, the proposed algorithm is able to match the FER performance of the state-of-the art fast CA-SCL-based decoders with up to $L=16$, while keeping an average computational complexity similar to that of a single SC decoder. Synthesis results in TSMC $65$nm CMOS technology shows that the proposed Fast-DSCF decoders is able to achieve an average throughput of up to $1.25$ Gbps, while being up to $5.8\times$ more area-efficient compared to the fast CA-SCL decoders with equivalent FER performance. Finally, observations at the increase trends in energy consumption with improved performance indicate that the Fast-DSCF is more energy-efficient than its CA-SCL-based counterparts. 

The structure of this paper is as follows: Preliminaries are described in Section~\ref{sec:bg}. Approximations to replace the transcendental computations in DSCF decoding are explained in Section~\ref{sec:metricapprox}. Fast decoding techniques for DSCF decoding are detailed in Section~\ref{sec:fastdscf}. Reducing the computational effort for fast decoding techniques is discussed in Section~\ref{sec:reducedsearch}. The hardware architecture for the Fast-DSCF decoding is explained in Section~\ref{sec:arch}, followed by simulation and implementation results in Section~\ref{sec:results}. Conclusions are drawn in Section~\ref{sec:conclusion}. Note that, a portion of this study has been discussed previously in \cite{DSCF-ICASSP20}.

\vspace{-1.25em}

\section{Preliminaries}\label{sec:bg}

Vectors and matrices are denoted with bold letters ($\boldsymbol{v}$), an index of a vector is denoted with a subscript ($\boldsymbol{v}_i$), a range of indices from $i$ to $j$ for a vector is denoted as $\boldsymbol{v}_{i:j}$. For LLRs ($L$) and partial sums ($\beta$) at decoding tree stage $S$ are denoted using a superscript ($L^{S}$, $\beta^S$). $\mathcal{E}_{\omega}$ denote the set of bit-flipping indices for the DSCF algorithm, and $L^{S}[\mathcal{E}_{\omega}]_{i}$ denote the LLR at stage $S$ and at index $i$ when the decisions at indices in $\mathcal{E}_{\omega}$ are flipped.

\subsection{Polar Codes}\label{sec:bg:sc}

A polar code $PC(N,K)$ splits $N$ channels into $K$ reliable ones that are used to transmit the information bits, and $N-K$ unreliable ones, which are frozen to a known value (usually to $0$). The set of frozen and non-frozen indices are denoted with $\mathcal{A}^C$ and $\mathcal{A}$, respectively. The encoding of a polar code is a linear transformation, such that $\boldsymbol{x} = \boldsymbol{u}\boldsymbol{G}^{\otimes n}\text{,}$ where $\boldsymbol{x}$ is the encoded vector, $\boldsymbol{u}$ is the message vector, and the generator matrix $\boldsymbol{G}^{\otimes n}$ is the $n$-th Kronecker product ($\otimes$) of the polar code kernel $\boldsymbol{G} =  \left[\begin{smallmatrix} 1&0\\ 1&1 \end{smallmatrix} \right]$ and $n = \log_2 N$, $n \in \mathbb Z^+$.

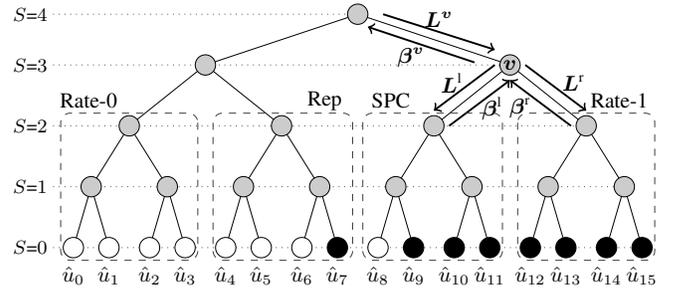
\begin{figure}
  \centering
  \scalebox{0.90}{
  %!TEX root = ../FastDSCF.tex
\begin{tikzpicture}[scale=.75]
\usetikzlibrary{backgrounds}

\filldraw[fill=gray!40!white, draw=black] (+0.00,+0.00) circle [radius=.2];

\filldraw[fill=gray!40!white, draw=black] (-3.00,-1.00) circle [radius=.2];
\filldraw[fill=gray!40!white, draw=black] (+3.00,-1.00) circle [radius=.2];

\filldraw[fill=gray!40!white, draw=black] (-4.50,-2.20) circle [radius=.2];
\filldraw[fill=gray!40!white, draw=black] (-1.50,-2.20) circle [radius=.2];
\filldraw[fill=gray!40!white, draw=black] (+1.50,-2.20) circle [radius=.2];
\filldraw[fill=gray!40!white, draw=black] (+4.50,-2.20) circle [radius=.2];

\filldraw[fill=gray!40!white, draw=black] (-5.25,-3.40) circle [radius=.2];
\filldraw[fill=gray!40!white, draw=black] (-3.75,-3.40) circle [radius=.2];
\filldraw[fill=gray!40!white, draw=black] (-2.25,-3.40) circle [radius=.2];
\filldraw[fill=gray!40!white, draw=black] (-0.75,-3.40) circle [radius=.2];
\filldraw[fill=gray!40!white, draw=black] (+0.75,-3.40) circle [radius=.2];
\filldraw[fill=gray!40!white, draw=black] (+2.25,-3.40) circle [radius=.2];
\filldraw[fill=gray!40!white, draw=black] (+3.75,-3.40) circle [radius=.2];
\filldraw[fill=gray!40!white, draw=black] (+5.25,-3.40) circle [radius=.2];

\filldraw[fill=white!40!white, draw=black] (-5.60,-4.60) circle [radius=.2];
\filldraw[fill=white!40!white, draw=black] (-4.90,-4.60) circle [radius=.2];
\filldraw[fill=white!40!white, draw=black] (-4.10,-4.60) circle [radius=.2];
\filldraw[fill=white!40!white, draw=black] (-3.40,-4.60) circle [radius=.2];
\filldraw[fill=white!40!white, draw=black] (-2.60,-4.60) circle [radius=.2];
\filldraw[fill=white!40!white, draw=black] (-1.90,-4.60) circle [radius=.2];
\filldraw[fill=white!40!white, draw=black] (-1.10,-4.60) circle [radius=.2];
\filldraw[fill=black!40!black, draw=black] (-0.40,-4.60) circle [radius=.2];
\filldraw[fill=white!40!white, draw=black] (+0.40,-4.60) circle [radius=.2];
\filldraw[fill=black!40!black, draw=black] (+1.10,-4.60) circle [radius=.2];
\filldraw[fill=black!40!black, draw=black] (+1.90,-4.60) circle [radius=.2];
\filldraw[fill=black!40!black, draw=black] (+2.60,-4.60) circle [radius=.2];
\filldraw[fill=black!40!black, draw=black] (+3.40,-4.60) circle [radius=.2];
\filldraw[fill=black!40!black, draw=black] (+4.10,-4.60) circle [radius=.2];
\filldraw[fill=black!40!black, draw=black] (+4.90,-4.60) circle [radius=.2];
\filldraw[fill=black!40!black, draw=black] (+5.60,-4.60) circle [radius=.2];

\node [color=black] at (-5.60,-5.15) {\small $\hat{u}_0$};
\node [color=black] at (-4.90,-5.15) {\small $\hat{u}_1$};
\node [color=black] at (-4.10,-5.15) {\small $\hat{u}_2$};
\node [color=black] at (-3.40,-5.15) {\small $\hat{u}_3$};
\node [color=black] at (-2.60,-5.15) {\small $\hat{u}_4$};
\node [color=black] at (-1.90,-5.15) {\small $\hat{u}_5$};
\node [color=black] at (-1.10,-5.15) {\small $\hat{u}_6$};
\node [color=black] at (-0.40,-5.15) {\small $\hat{u}_7$};
\node [color=black] at (+0.40,-5.15) {\small $\hat{u}_8$};
\node [color=black] at (+1.10,-5.15) {\small $\hat{u}_9$};
\node [color=black] at (+1.90,-5.15) {\small $\hat{u}_{10}$};
\node [color=black] at (+2.60,-5.15) {\small $\hat{u}_{11}$};
\node [color=black] at (+3.40,-5.15) {\small $\hat{u}_{12}$};
\node [color=black] at (+4.10,-5.15) {\small $\hat{u}_{13}$};
\node [color=black] at (+4.90,-5.15) {\small $\hat{u}_{14}$};
\node [color=black] at (+5.60,-5.15) {\small $\hat{u}_{15}$};

\begin{scope}[on background layer]
\draw [-] (+0.00,+0.00) -- (-3.00,-1.00);
\draw [-] (+0.00,+0.00) -- (+3.00,-1.00);

\draw [-] (-3.00,-1.00) -- (-4.50,-2.20);
\draw [-] (-3.00,-1.00) -- (-1.50,-2.20);
\draw [-] (+3.00,-1.00) -- (+1.50,-2.20);
\draw [-] (+3.00,-1.00) -- (+4.50,-2.20);

\draw [-] (-4.50,-2.20) -- (-5.25,-3.40);
\draw [-] (-4.50,-2.20) -- (-3.75,-3.40);
\draw [-] (-1.50,-2.20) -- (-2.25,-3.40);
\draw [-] (-1.50,-2.20) -- (-0.75,-3.40);
\draw [-] (+1.50,-2.20) -- (+0.75,-3.40);
\draw [-] (+1.50,-2.20) -- (+2.25,-3.40);
\draw [-] (+4.50,-2.20) -- (+3.75,-3.40);
\draw [-] (+4.50,-2.20) -- (+5.25,-3.40);

\draw [-] (-5.25,-3.40) -- (-5.60,-4.60);
\draw [-] (-5.25,-3.40) -- (-4.90,-4.60);
\draw [-] (-3.75,-3.40) -- (-4.10,-4.60);
\draw [-] (-3.75,-3.40) -- (-3.40,-4.60);
\draw [-] (-2.25,-3.40) -- (-2.60,-4.60);
\draw [-] (-2.25,-3.40) -- (-1.90,-4.60);
\draw [-] (-0.75,-3.40) -- (-1.10,-4.60);
\draw [-] (-0.75,-3.40) -- (-0.40,-4.60);
\draw [-] (+0.75,-3.40) -- (+0.40,-4.60);
\draw [-] (+0.75,-3.40) -- (+1.10,-4.60);
\draw [-] (+2.25,-3.40) -- (+1.90,-4.60);
\draw [-] (+2.25,-3.40) -- (+2.60,-4.60);
\draw [-] (+3.75,-3.40) -- (+3.40,-4.60);
\draw [-] (+3.75,-3.40) -- (+4.10,-4.60);
\draw [-] (+5.25,-3.40) -- (+4.90,-4.60);
\draw [-] (+5.25,-3.40) -- (+5.60,-4.60);

\draw [-,dotted,color=white!30!black] (-6.00,+0.00) -- (+0.00,+0.00);
\draw [-,dotted,color=white!30!black] (-6.00,-1.00) -- (+3.00,-1.00);
\draw [-,dotted,color=white!30!black] (-6.00,-2.20) -- (+4.50,-2.20);
\draw [-,dotted,color=white!30!black] (-6.00,-3.40) -- (+5.25,-3.40);
\draw [-,dotted,color=white!30!black] (-6.00,-4.60) -- (+5.60,-4.60);

\end{scope}

\node [color=black] at (-6.45,+0.00) {\footnotesize $S$=$4$};
\node [color=black] at (-6.45,-1.00) {\footnotesize $S$=$3$};
\node [color=black] at (-6.45,-2.20) {\footnotesize $S$=$2$};
\node [color=black] at (-6.45,-3.40) {\footnotesize $S$=$1$};
\node [color=black] at (-6.45,-4.60) {\footnotesize $S$=$0$};

%SHOWING ALPHA AND BETA TRANSACTIONS OVER SC TREE:
\node [color=black] at (+3.00,-1.00) {\small $\boldsymbol{v}$};

\draw [->,thick] (+0.50,+0.00) -- (+2.70,-0.70);
\node [color=black,rotate=0] at (+1.60,-0.10) {\small $\boldsymbol{\bm{L}^{v}}$};
\draw [<-,thick] (+0.20,-0.25) -- (+2.30,-0.95);
\node [color=black,rotate=0] at (+1.05,-0.85) {\small $\boldsymbol{\bm{\beta}^{v}}$};

\draw [->,thick] (+2.70,-1.00) -- (+1.50,-1.90);
\node [color=black] at (+1.85,-1.35) {\small $\boldsymbol{\bm{L}^{\text{l}}}$};
\draw [->,thick] (+3.30,-1.00) -- (+4.50,-1.90);
\node [color=black] at (+4.25,-1.35) {\small $\boldsymbol{\bm{L}^{\text{r}}}$};

\draw [<-,thick] (+3.00,-1.30) -- (+1.80,-2.20);
\node [color=black] at (+2.65,-1.85) {\small $\boldsymbol{\beta^{\text{l}}}$};
\draw [<-,thick] (+3.00,-1.30) -- (+4.20,-2.20);
\node [color=black] at (+3.20,-1.85) {\small $\boldsymbol{\beta^{\text{r}}}$};

\draw [rounded corners, dashed, color=darkgray](+3.15,-1.95) rectangle (+5.85,-4.85);	%R1 NODE
\draw [rounded corners, dashed, color=darkgray](+0.10,-1.95) rectangle (+2.85,-4.85);	%SPC-4 NODE
\draw [rounded corners, dashed, color=darkgray](-3.15,-1.95) rectangle (-5.85,-4.85);	%R0 NODE
\draw [rounded corners, dashed, color=darkgray](-0.10,-1.95) rectangle (-2.85,-4.85);	%SPC-4 NODE

\node [color=black] at (-5.30,-1.70) {\small Rate-0};
\node [color=black] at (-0.65,-1.70) {\small Rep};
\node [color=black] at (+0.65,-1.70) {\small SPC};
\node [color=black] at (+5.15,-1.70) {\small Rate-1};

\end{tikzpicture}}
  \caption{Successive cancellation decoding tree for $PC(16,8)$. LLR ($\bm{L}$) and partial sum ($\beta$) vectors of parent node $v$ and of child nodes are represented with superscripts that indicate the direction ($l$ for left, $r$ for right) and not with their in-text superscripts for simplicity. Stages ($S$) for each level and the sub-codes with special frozen bit-patterns (Rate-0, Rate-1, Rep, SPC) are outlined for reference.}
  \label{fig:scdecode_n16}
\end{figure}

The decoding schedule of SC can be interpreted as a binary tree search that starts from the root node (that contains the channel observation), and with priority given to the left branch. 
An illustration of the SC decoder tree is shown in Fig.~\ref{fig:scdecode_n16} for $PC(16,8)$.
Each stage in the tree is defined by the inverse of its depth from the root node, which is denoted by $S$ where $0 \leq S \leq n$. Each node contains $N_v = 2^S$ soft information, interpreted in log-likelihood ratio (LLR) form ($\bm{L^{S}}$) that are propagated to their child nodes. In return, each child node propagates $N_v$ hard information ($\bm{\beta^{S}}$) to their parent nodes, called partial sums. %In Fig.~\ref{fig:scdecode_n16}, the superscripts used for the LLR ($\bm{L}$) and partial sum ($\bm{\beta}$) vectors indicate the parent node $v$, and direction ($\text{l}$ for left, $\text{r}$ for right) for child nodes instead of their in-text superscripts for simplicity.
As illustrated in Fig.~\ref{fig:scdecode_n16}, from a node $v$ that has $\boldsymbol{L}^{v}$ LLRs, the LLRs at the left child ($\boldsymbol{L}^{\text{l}}$) and the right child ($\boldsymbol{L}^{\text{r}}$) are calculated as
% \begin{equation}\label{eqn:alphaleft}
% L^{S-1}_i = \sgn(L^{S}_{i})\sgn(L^{S}_{i+2^{S-1}}) \min(|L^{S}_{i}|,|L^{S}_{i+2^{S-1}}|) \text{,} 
% \end{equation}
% and the right child as
% \begin{equation}\label{eqn:alpharight}
% L^{S-1}_i = L^{S}_{i+2^{S-1}} + (1-2\beta^\text{left}_{i})L^{S}_{i} \text{,} 
% \end{equation}
\begin{equation}\label{eqn:alphaleft}
L^{\text{l}}_i = \sgn(L^{v}_{i})\sgn(L^{v}_{i+2^{S-1}}) \min(|L^{v}_{i}|,|L^{v}_{i+2^{S-1}}|) \text{,} 
\end{equation}
\begin{equation}\label{eqn:alpharight}
L^{\text{r}}_i = L^{v}_{i+2^{S-1}} + (1-2\beta^\text{l}_{i})L^{v}_{i} \text{.} 
\end{equation}
Given that the hard decision information from the left child ($\boldsymbol{\beta}^{\text{l}}$) and the right child ($\boldsymbol{\beta}	^{\text{r}}$) of node $v$ are available, the $\boldsymbol{\beta}^{v}$ for node $v$ is calculated as
\begin{equation}\label{eqn:beta}
  \beta_i^v=\left\{
  \begin{array}{@{}ll@{}}
    \beta^{\text{l}}_{i} \oplus \beta^{\text{r}}_{i}, & \text{if}~ i \leq 2^{S-1} \\
    \beta^{\text{r}}_{i-2^{S-1}}, & \text{otherwise.}
  \end{array}\right.
\end{equation}

% and at decoding stages $S<n$ the hard decision information ($\bm{\beta^S}$) are propagated from the left ($\beta^{\text{left}}$) and right ($\beta^{\text{right}}$) children as
% \begin{equation}\label{eqn:beta}
%   \beta_i^S=\left\{
%   \begin{array}{@{}ll@{}}
%     \beta^{\text{left}}_{i} \oplus \beta^{\text{right}}_{i}, & \text{if}~ i \leq 2^{S-1} \\
%     \beta^{\text{right}}_{i-2^{S-1}}, & \text{otherwise.}
%   \end{array}\right.
% \end{equation}
The bit estimations are performed at leaf node stage $S=0$ sequentially, starting from the leftmost index. Estimation of each bit $\hat{u}_i$ depends on the channel observation $\boldsymbol{y}$ and previously decoded bits $\hat{\boldsymbol{u}}_{0:i-1}$, such that
\begin{equation}\label{eqn:scdecode}
\hat{u}_i=\left\{
  \begin{array}{@{}ll@{}}
    0, & \text{if } \text{Pr}[\boldsymbol{y},\hat{\boldsymbol{u}}_{0:i\text{-}1} | u_i = 0] \geq  \text{Pr}[\boldsymbol{y},\hat{\boldsymbol{u}}_{0:i\text{-}1} | u_i = 1]; \\
    0, & \text{if } i \in \mathcal{A}^C; \\
    1, & \text{otherwise.}
  \end{array}\right.
\end{equation}

It was shown in \cite{SSC2011} and \cite{sarkis14} that nodes in the SC decoding tree with special frozen bit patterns are not needed to be explicitly traversed; dedicated fast decoding techniques for such special nodes improves the throughput of the decoding substantially. Among these special nodes, decoding of Rate-0 (where all indices are frozen) Rate-1 (where no indices are frozen), repetition (Rep, where only the rightmost index is non-frozen) and single-parity check (SPC, where only the leftmost index is frozen) are within the scope of this work. An example for each of the considered nodes are highlighted with the dashed lines in Fig.~\ref{fig:scdecode_n16}.

\vspace{-1em}	

\subsection{SC-Flip and Dynamic SC-Flip Decoding}\label{sec:bg:scf}

In a failed SC decoding, the incorrect bit estimations (\textit{e.g.} errors) could occur in two different ways. The first way is due to the the noise present in the channel, this type of errors is called a \textit{channel-induced error}. The second way to incur an error is due to a previously made error during the sequential schedule of SC decoding (\ref{eqn:scdecode}), which we call a \textit{propagated error}. Since the first error in the codeword cannot be propagated from a previously made error, it is a channel-induced error. Therefore, if this error is found and corrected, then its associated propagated errors -- if there are any -- also disappear. In this context, while propagated errors are dependent on their associated channel errors, the channel errors are independent from one another.

The observation above was originally made in \cite{SCFlip14}, and it was also observed that most of the decoding failures are due to a single channel-induced error. Hence, if a single channel-induced error was avoided, then the error-correction performance would improve. Aided by an outer cyclic redundancy check (CRC) code for detecting whether an initial SC decoding has failed, SCF decoding first creates a list of bit-flipping positions using non-frozen leaf indices sorted according to their LLR magnitudes. Then, the SC decoding process is relaunched but the hard decision at the index that holds the next lowest LLR magnitude is flipped. This is repeated until no errors are detected or until all positions in the list have been considered. When SCF decoding fails, it is either due to: (i) a wrong codeword with a valid CRC, or (ii) not locating the correct erroneous index within a maximum number of additional attempts ($T_{\text{max}}$), or (iii) having more than one channel-induced error in the codeword.

The performance improvement in error-correction introduced by the SCF decoding algorithm is limited due to two different problems. The first problem is that the SCF algorithm cannot correct more than a single channel-induced error. Even though several attempts lead to improvements in the SCF decoding \cite{SCF-WCNC18,SCFlip_TCOM18,SCF_Kim_2018,SCF_Lv_2019,FTSCF_ICC20}, they are unable to tackle more than one channel-induced error.
An alternative approach that segments the codeword into multiple partitions, and applying SCF decoding to each partition separately is also shown to have limited performance improvements \cite{PSCF-ICC18,PSCF_IEEEAccess2019_Li,PSCF_SpliTech2019}.
The second problem is that the decision metric used in the SCF decoding is not able to distinguish channel-induced errors from propagated errors. Indeed, we have observed that the propagated errors may also carry small LLR magnitudes \cite{SCFlip_TCOM18} which is the only parameter that the SCF decoding relies on for identifying channel-induced errors. 

The proposals that address the limited performance improvement problem of SCF decoding in the literature can be classified into two different techniques. The first technique is to merge the SCF algorithm with the SCL algorithm \cite{SCLF_Yongrun_2018,SCLF_Cheng_2019,SCLF_Cao_2019}. This approach has shown to improve the decoding performance at the cost of an increased computational complexity. The second technique is to create combinations of bit-flipping positions to tackle more than one channel induced error \cite{SCFlip17-conf,SCFlip17-jour,SCF-GLOBECOM17,ProgressiveBitFlip_IEEEAccess2018,ProgressiveSCF_Cui_2018,SCF_Wang_2019,Carlo_SCFlip_ITW2019}. Among these, the Dynamic SCF (DSCF) algorithm \cite{SCFlip17-jour} tackles both of the problems associated with the SCF decoding simultaneously: (i) the search for bit-flipping is not limited to a single channel-induced error, and (ii) the decision metric is more efficient in identifying the correct bit-flipping positions than the SCF algorithm. 

To identify more than one channel-induced error, DSCF updates the set of flipping indices progressively over the course of each decoding attempt: Let $\mathcal{E}_{\omega} = \{i_1, \dots, i_{\omega}\}$ denote the set of bit-flipping indices at an additional decoding attempt, where $i_1 < \cdots < i_{\omega}$ and $0 \leq \omega \leq K+C$. Note that here, $C$ is the CRC remainder length. In this sense, $\omega$ is the number of attempted channel-induced errors, which is referred to as \textit{the decoding order}. $\mathcal{E}_{\omega}$ is built progressively over a prior additional decoding attempt with $\mathcal{E}_{\omega-1} = \{i_1, \dots, i_{\omega-1}\}$.

Unlike in the SCF decoding, the decision metric of DSCF decoding for non-frozen indices does not only depend on their LLR magnitudes. Instead, all the decisions that were made at the prior non-frozen indices are also considered to calculate a more comprehensive decision metric. In this sense, let $\text{Pr}(\mathcal{E}_{\omega})$ be the probability of SC decoding being successful after flipping the bits in $\mathcal{E}_{\omega}$. It was shown in \cite{SCFlip17-jour} that $\text{Pr}(\mathcal{E}_{\omega})$ can be formulated as 
\begin{equation}\label{eqn:prob_dscf}
	\text{Pr}(\mathcal{E}_{\omega}) = \prod_{j \in \mathcal{E}_{\omega}} p_e(\hat{u}[\mathcal{E}_{\omega-1}]_j) \times \prod_{\substack{j < i_{\omega} \\ j \in \mathcal{A} \setminus \mathcal{E}_{\omega}}} \big{(}1-p_e(\hat{u}[\mathcal{E}_{\omega-1}]_j)\big{)}
\end{equation}
where $p_e(\hat{u}[\mathcal{E}_{\omega-1}]_j)$ is the probability of incurring an error at index $j$, such that
\begin{align}\label{eqn:perr}
	p_e(\hat{u}[\mathcal{E}_{\omega-1}]_j) \coloneqq  \text{Pr}(& \hat{u}[\mathcal{E}_{\omega-1}]_j \neq u_j|\boldsymbol{y}, \notag\\ & \hat{\boldsymbol{u}}[\mathcal{E}_{\omega-1}]_{0:j-1} = \boldsymbol{u}_{0:j-1} )\text{.}
\end{align}

Let us elaborate on the computation of (\ref{eqn:prob_dscf}) with a simple example. Assume that the estimations $\hat{u}_{7}$ and $\hat{u}_{12}$ in the $PC(16,8)$ polar code in Fig.~\ref{fig:scdecode_n16} have \textit{channel-induced errors}. Let a specific $\mathcal{E}_{\omega}$ include the erroneous indices, \textit{i.e.} $\omega=2$ and $\mathcal{E}_{2} = \{7,12\}$. By the successive course of the decoding, $\hat{u}_{7}$ is flipped first. The probability of a bit-flip at index $7$ yielding the correct decision is equal to the probability of index $7$ incurring a channel-induced error originally. %, which is $p_e(\hat{u}[\mathcal{E}_{1}]_{i_{7}})$. 
The bit estimations that follow (i.e. at indices $9$, $10$, $11$) are impacted by the first bit-flip and therefore denoted as $\hat{u}[\mathcal{E}_{1}]_j$. Consequently, their associated probability of correct estimation are represented as $\big{(}1-p_e(\hat{u}[\mathcal{E}_{1}]_j)\big{)}$. Finally, the probability of error at the last index of $\mathcal{E}_{2}$ (which is index $12$), is the same as the probability of incurring an error when it is not corrected, which is $p_e(\hat{u}[\mathcal{E}_{1}]_{12})$. The product of all these probabilities creates the probability of the successful decoding after flipping the indices of $\mathcal{E}_{\omega}$, which is summarized in (\ref{eqn:prob_dscf}).

It can be seen in (\ref{eqn:perr}) that $p_e(\hat{u}[\mathcal{E}_{\omega-1}]_j)$ depends on all the previous bits decoded correctly, which cannot be granted in practice. Hence, an approximation to $p_e(\hat{u}[\mathcal{E}_{\omega-1}]_j)$, that depends on all the previously decoded bits, regardless of them being correctly decoded, can be used instead:
\begin{equation}\label{eqn:perrapprox}
	q_e(\hat{u}[\mathcal{E}_{\omega-1}]_j) = \frac{1}{1+\exp(|L^{0}[\mathcal{E}_{\omega-1}]_j|)} , \forall j \in \mathcal{A}
\end{equation}
where $L^{0}[\mathcal{E}_{\omega-1}]_j$ is the LLR at index $j$ of the current decoding attempt. Hence, approximating (\ref{eqn:perr}) with (\ref{eqn:perrapprox}) and thus substituting (\ref{eqn:perrapprox}) into (\ref{eqn:prob_dscf}) yields the decision metric $m$ for the bit-flipping set $\mathcal{E}_{\omega}$: 
\begin{align}\label{eqn:dscfmetricMnonlog}
m(\mathcal{E}_{\omega}) &~ =  \prod_{j \in \mathcal{E}_{\omega}} \frac{1}{1+\exp(|L^{0}[\mathcal{E}_{\omega-1}]_j|)} \notag\\ & \times \prod_{\substack{j < i_{\omega} \\ j \in \mathcal{A} \setminus \mathcal{E}_{\omega}}} \big{(}1-\frac{1}{1+\exp(|L^{0}[\mathcal{E}_{\omega-1}]_j|)}\big{)}
\end{align}
Using the fact that $\frac{1}{1+\exp(x)} = \frac{\exp(-x)}{1+\exp(-x)}$, (\ref{eqn:dscfmetricMnonlog}) can also be written as 
\begin{align*}\label{eqn:dscfmetricMnonlog2}
m(\mathcal{E}_{\omega}) &~ =  \prod_{j \in \mathcal{E}_{\omega}} \exp(-|L^{0}[\mathcal{E}_{\omega-1}]_j|) \notag\\ & \times \prod_{\substack{j \leq i_{\omega} \\ j \in \mathcal{A} }} \frac{1}{1+\exp(|-L^{0}[\mathcal{E}_{\omega-1}]_j|)}
\end{align*}

For numerical stability, the metric is converted to the logarithmic domain using $M(\mathcal{E}_{\omega}) = -\log(m(\mathcal{E}_{\omega}))$ as
\begin{equation}\label{eqn:dscfmetricMbad}
	M(\mathcal{E}_{\omega})  =  \sum_{j \in \mathcal{E}_{\omega}} | L^{0}[\mathcal{E}_{\omega-1}]_j | + S(\mathcal{E}_{\omega}),
\end{equation}
where
\begin{equation}\label{eqn:dscfmetricSbad}
	S(\mathcal{E}_{\omega})  = \sum_{\substack{{j \leq i_{\omega}}\\ j \in \mathcal{A}}} \log(1+\exp(-| L^{0}[\mathcal{E}_{\omega-1}]_j | ) )\text{.}
\end{equation}

Finally, in order to approximate the value of $q_e(\hat{u}[\mathcal{E}_{\omega-1}]_j)$ (\ref{eqn:perrapprox}) close to $p_e(\hat{u}[\mathcal{E}_{\omega-1}]_j)$ (\ref{eqn:perr}), a \textit{perturbation parameter} $\alpha$ was defined in \cite{SCFlip17-jour} and used, such that
\begin{equation}\label{eqn:dscfmetricM}
	M_{\alpha}(\mathcal{E}_{\omega})  =  \sum_{j \in \mathcal{E}_{\omega}} | L^{0}[\mathcal{E}_{\omega-1}]_j | + S_{\alpha}(\mathcal{E}_{\omega}),
\end{equation}
and 
\begin{equation}\label{eqn:dscfmetricS}
	S_{\alpha}(\mathcal{E}_{\omega})  = \frac{1}{\alpha} \sum_{\substack{{j \leq i_{\omega}}\\ j \in \mathcal{A}}} \log(1+\exp(-\alpha | L^{0}[\mathcal{E}_{\omega-1}]_j | ) )\text{.}
\end{equation}
The value of $\alpha$ can be optimized via Monte-Carlo simulations \cite{SCFlip17-jour} or machine learning \cite{Doan_NDSCF}.

\begin{algorithm}[t]
\SetKwData{Left}{left}\SetKwData{This}{this}\SetKwData{Up}{up}
\SetKwFunction{Union}{Union}
\SetKwFunction{back}{back}
\SetKwFunction{pop}{pop}
\SetKwFunction{push}{push}
\SetKwFunction{sort}{sort}
\SetKwInOut{Input}{input}\SetKwInOut{Output}{output}
  \textbf{procedure} DSCF$(\boldsymbol{y}_{0:N-1},T_{\text{max}}, \omega,\mathcal{A})$ \\
  % initialize: $t \leftarrow 0$, $w \leftarrow 0$ \\
  initialize: $t \leftarrow 0$, $w \leftarrow 0$, $\mathcal{E}_{w} \leftarrow \varnothing$, $\boldsymbol{V} = \{v^{w} \leftarrow w; v^{\text{idx}} \leftarrow \varnothing; v^{M_{\alpha}} \leftarrow +\infty \}$. \\
  $(\hat{\boldsymbol{u}}_{0:N-1}, L^{0}_{0:N-1}) \leftarrow \text{SCF}(\varnothing)$ \\
  \If{$\big{(}!\text{CRC}(\hat{\boldsymbol{u}}_{0:N-1},\mathcal{A}) ~\&~ T_{\text{max}} > 0  ~\&~ \omega > 0 \big{)} $}{
		$\boldsymbol{V}^{\text{new}} \leftarrow \varnothing$ \\
		\For{$i \in \mathcal{A} $}{
			\text{compute} $M_{\alpha}(\mathcal{E}_{1})$ (Eq. (\ref{eqn:dscfmetricM})) \\
			%$M_{\alpha}(\mathcal{E}_{1}) \leftarrow L^{0}_{i}$ (Eq. (\ref{eqn:dscfmetricM})) \\
			$\boldsymbol{V}^{\text{new}} \leftarrow \push(\{1; \{i\}; M_{\alpha}(\mathcal{E}_{1})\}) $\\
		}
	 	$\boldsymbol{V} \leftarrow \sort(\boldsymbol{V}^{\text{new}} \rightarrow v^{M_{\alpha}} , \text{ascending})$ \\
	 \While{$(!\text{CRC}(\hat{\boldsymbol{u}}_{0:N-1},\mathcal{A}) ~\&~ t < T_{\text{max}} )$ }{% ~\&~ w \leq \omega)$ }{
	 	$t \leftarrow t+1$ \\
	 	$w \leftarrow V_{0}(v^{\text{w}})$ \\
	 	$\mathcal{E}_{w} \leftarrow V_{0}(v^{\text{idx}})$ \\
	  	$\boldsymbol{V} \leftarrow \pop(V_{0})$ \\
	  	$(\hat{\boldsymbol{u}}_{0:N-1}, L^{0}_{0:N-1}) \leftarrow \text{SCF}(\mathcal{E}_{w})$ \\
	 	\If{$w < \omega$}{
	 	$\boldsymbol{V}^{\text{new}} \leftarrow \varnothing$ \\
		\For{$i > \back(\mathcal{E}_{w}), i \in \mathcal{A} $}{
			\text{compute} $M_{\alpha}(\mathcal{E}_{w+1})$ (Eq. (\ref{eqn:dscfmetricM})) \\
			%$M_{\alpha}(\mathcal{E}_{w}) \leftarrow L^{0}_{i}$ (Eq. (\ref{eqn:dscfmetricM})) \\
			$V^{\text{new}} \leftarrow \push(\{w+1,; \mathcal{E}_{w} \cup \{i\}; M_{\alpha}(\mathcal{E}_{w+1})\}) $\\
		}
		$\boldsymbol{V} \leftarrow \sort(\boldsymbol{V}^{\text{new}} \cup \boldsymbol{V} \rightarrow v^{M_{\alpha}} , \text{ascending})$ \\
		}

	  }
  }
  \textbf{return} $\hat{\boldsymbol{u}}_{0:N-1}$ \\
  \caption{Dynamic SCF Algorithm}\label{alg:dscf}
\end{algorithm}

The procedure of the DSCF decoding is summarized in Algorithm~\ref{alg:dscf}. Required inputs are the channel LLRs, maximum number of iterations $T_{\text{max}}$, maximum bit-flipping order $\omega$ and $\mathcal{A}$. The received codeword is initially decoded with the SC algorithm (line 3). Here, the input for SCF algorithm is the bit-flipping indices, hence SCF with an empty input is equivalent to SC decoding. If the CRC on the estimated information bits fails and if there is room for additional bit-flips (line 4), then an initial list of bit-flipping indices are created using leaf LLRs (lines 5-8). Here, a special data structure ($V$) is used to enclose the bit-flipping order ($v^{w}$), bit-flipping indices ($v^{\text{idx}}$) and the metric ($v^{M_{\alpha}}$). The created vector of $V$ is then sorted with respect to their metric value (indicated with $\rightarrow$) in ascending order (line 9). Following, a series of decoding iterations is initiated that is conditioned on the CRC output and $T_{\text{max}}$ (lines 10-21). At each iteration, the decoding order and the bit-flipping indices are obtained from the next item in $V$ (lines 12-13). The used entry from $V$ is discarded from the list ($\texttt{pop}()$ in line 14). The indices are used as an input to the new SCF decoding attempt (line 15). If the new decoding attempt allows for further bit-flipping investigations over the newly created decoding trajectory (line 16), then new bit-flipping indices are built on top of the current bit-flipping attempt (note the $\mathcal{E}_{w} \cup \{i\}$ in line 20), and the updated vector is re-sorted before the next iteration begins. Here, the evaluated bit-flipping indices must be greater than the last index at $\mathcal{E}_{w}$ (obtained by $\texttt{back}()$ operation at line 18), following the definition of $\mathcal{E}_{w}$. If the initial SC decoding has a valid CRC, or when at least one loop condition is broken, the bit estimation is reported (line 22).

To gain an in-depth theoretical background on the derivation of (\ref{eqn:dscfmetricM}) and the DSCF algorithm, the readers are strongly encouraged to refer to Section V-A and Section VI-A of \cite{SCFlip17-jour}, respectively.

\section{Approximations for DSCF Decoding}\label{sec:metricapprox}

\begin{figure}[t]
  \centering
  %!TEX root = ../FastDSCF.tex

\begin{tikzpicture}[spy using outlines=
	{circle, magnification=2.0, connect spies}]

    \begin{semilogyaxis}[
            footnotesize, width=\columnwidth, height=.60\columnwidth,    
            xmin=1, xmax=3.25, xtick={1, 1.25,...,3.25},
            % ymin=1e-7,  
            ymax=1,
            xlabel=SNR (dB),%$\frac{E_s}{N_0} \text{~(dB)}$, %xlabel style={yshift=0.4em},
            ylabel=FER,  %ylabel style={yshift=-0.75em},
            grid=both, grid style={gray!30},
            tick align=outside, tickpos=left, 
            legend cell align={left},
            legend columns=4,
            legend pos=south west,
            legend image code/.code={
                \draw[mark repeat=2,mark phase=2]
                plot coordinates {
                (0cm,0cm)
                (0.15cm,0cm)        %% default is (0.3cm,0cm)
                (0.3cm,0cm)         %% default is (0.6cm,0cm)
                };%
                },
            legend style={font=\scriptsize},
        ]

\addlegendimage{empty legend}
\addlegendentry{SCF}

%w=1,t=10
\addplot[
    color=Paired-5,
    mark=o,
    semithick
]
table {
1.00 7.42000e-01
1.25 5.32200e-01
1.50 3.21700e-01
1.75 1.57700e-01
2.00 6.15000e-02
2.25 2.21000e-02
2.50 6.40000e-03
2.75 1.87822e-03
3.00 3.44372e-04
3.25 8.13373e-05
};
\addlegendentry{$w=1$}

%w=2,t=40
\addplot[
    color=Paired-5,
    mark=x,
    semithick,
]
table {
1.00 7.19700e-01
1.25 5.08100e-01
1.50 2.99800e-01
1.75 1.42000e-01
2.00 5.41000e-02
2.25 2.00000e-02
2.50 5.50000e-03
2.75 1.66639e-03
3.00 3.23989e-04
3.25 5.09490e-05
};
\addlegendentry{$w=2$}

%w=3,t=200
\addplot[
    color=Paired-5,
    mark=triangle,
    semithick
]
table {
1.00 6.89800e-01
1.25 4.69500e-01
1.50 2.66800e-01
1.75 1.21300e-01
2.00 4.38000e-02
2.25 1.51000e-02
2.50 3.72162e-03
2.75 1.01767e-03
3.00 1.62653e-04
3.25 2.45024e-05
};
\addlegendentry{$w=3$}

\addlegendimage{empty legend}
\addlegendentry{DSCF}

\addplot[
    color=Paired-1,
    mark=o,
    semithick
]
table {
1.00e+00 7.42900e-01
1.25e+00 5.21900e-01
1.50e+00 2.82800e-01
1.75e+00 1.10100e-01
2.00e+00 3.38000e-02
2.25e+00 8.10000e-03
2.50e+00 1.68816e-03
2.75e+00 3.36945e-04
3.00e+00 7.92511e-05
3.25e+00 1.17657e-05
};
\addlegendentry{$w=1$}

\addplot[
    color=Paired-1,
    mark=x,
    semithick
]
table {
1.00 6.69900e-01
1.25 4.21600e-01
1.50 1.90200e-01
1.75 5.51000e-02
2.00 1.04000e-02
2.25 1.60591e-03
2.50 1.93149e-04
2.75 2.45751e-05
3.00 2.77045e-06
3.25 5.01661e-07
};
\addlegendentry{$w=2$}

\addplot[
    color=Paired-1,
    mark=triangle,
    semithick
]
table {
1.00 0.57909          
1.25 0.3125           
1.50 0.11543          
1.75 0.02673          
2.00 0.00409          
2.25 0.000395708          
2.50 0.0000268839            
2.75 0.00000261407  
3.00 3.60310e-07
3.25 6.14144e-08 
};
\addlegendentry{$w=3$}

\addlegendimage{empty legend}
\addlegendentry{SCO}

\addplot[
    color=Paired-7!50!Paired-8,
    dashed,
    mark options={solid},
    mark=o,
    semithick
]
table {
1.00 6.59430e-01
1.25 4.27870e-01
1.50 2.20750e-01
1.75 9.03500e-02
2.00 2.95000e-02
2.25 7.42000e-03
2.50 1.73000e-03
2.75 2.99287e-04
3.00 7.08938e-05
3.25 1.43373e-05
3.50 2.92422e-06
3.75 6.66162e-07
4.00 1.30563e-07
};
\addlegendentry{$w=1$}

\addplot[
    color=Paired-7!50!Paired-8,
    dashed,
    mark options={solid},
    mark=x,
    semithick
]
table {
1.00 4.76790e-01
1.25 2.50240e-01
1.50 9.71400e-02
1.75 2.83400e-02
2.00 6.06000e-03
2.25 1.06000e-03
2.50 1.41300e-04
2.75 1.64007e-05
3.00 1.92590e-06
3.25 4.18208e-07
};
\addlegendentry{$w=2$}

\addplot[
    color=Paired-7!50!Paired-8,
    dashed,
    mark options={solid},
    mark=triangle,
    semithick
]
table {
1.00 3.30310e-01
1.25 1.40990e-01
1.50 4.19600e-02
1.75 9.17000e-03
2.00 1.38000e-03
2.25 1.72022e-04
2.50 1.63814e-05
2.75 1.11383e-06
3.00 1.03196e-07
3.25 1.69226e-08
};
\addlegendentry{$w=3$}

\end{semilogyaxis}
\end{tikzpicture}
  \vspace{-2em}
  \caption{\label{fig:FER_SCFvsDSCF} Error-correction performance comparison of SCF and DSCF algorithms at $\omega \in \{1,2,3\}$ with $T_{\text{max}} \in \{10,50,200\}$, using $PC(1024,512)$ \cite{38.212} and $C=16$.}%0x1021, Tmax in 10,40,200.
\end{figure}
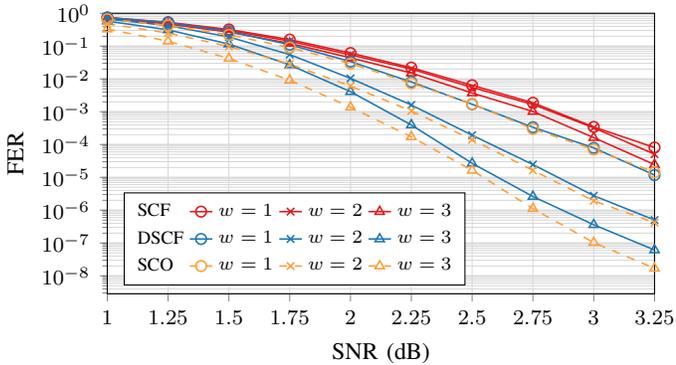

The $S_{\alpha}(\mathcal{E}_{\omega})$ component of the metric computation in DSCF decoding has a crucial impact on the identification of the correct bit-flipping indices; and thus, on the improvement of the error-correction performance. In fact, without $S_{\alpha}(\mathcal{E}_{\omega})$ the decision metric of DSCF reverts to the SCF decoding with higher error order. Fig.~\ref{fig:FER_SCFvsDSCF} illustrates the differences in FER performance using the metric of SCF and DSCF, at error orders $\omega \in \{1,2,3\}$ using $PC(1024,512)$ and $C=16$. 
For this exercise, the SCF algorithm is enhanced to tackle higher order errors by updating the set of flipping indices progressively similar to that of DSCF algorithm. In other words, SCF builds a bit-flipping set $\mathcal{E}_{\omega}$, however it uses only the LLR magnitude values of the flipping indices while building it. In this sense, SCF with $\omega=1$ is the original SCF decoding. 
For each error order, the dashed lines depict the ideal decoding performance when all $\omega$ errors are corrected using a genie-aided decoder, called SC-Oracle (SCO) \cite{SCFlip14}. Observe that even when the SCF is modified to correct higher order errors, the associated performance improvement is fractional such that the FER of SCF with $\omega=3$ is worse than that of DSCF with $\omega=1$. This implies that the metric computation of DSCF algorithm is more efficient, and essential for tackling higher-order errors. On the other hand, although DSCF has superior error-correction performance due to the $S_{\alpha}(\mathcal{E}_{\omega})$ term, its required logarithmic and exponential operations make DSCF inconvenient for efficient hardware implementations.

We reformulate $S_{\alpha}(\mathcal{E}_{\omega})$ in (\ref{eqn:dscfmetricS}) as 
\begin{equation}\label{eqn:S}
S_{\alpha}(\mathcal{E}_{\omega}) =  \sum_{\substack{{j \leq i_{\omega}}\\ j \in \mathcal{A}}} f_{\alpha}(| L^{0}[\mathcal{E}_{\omega-1}]_j |),
\end{equation}
where \vspace*{-.5em}
\begin{equation}\label{eqn:fx}
f_{\alpha}(x) = \frac{1}{\alpha} \log\big{(}1+\exp(-\alpha x)\big{)}.
\end{equation}

\begin{figure}[t]
  \centering
  %!TEX root = ../icassp2020_thibaud.tex

\begin{tikzpicture}

\begin{axis}[
	footnotesize,
    xlabel={$x$},
    grid=both,
    xmin=0,xmax=16,
    xtick={0,2, ..., 16},
    grid=both, grid style={gray!30},
    width=\columnwidth,
    height=0.50\columnwidth, 
    legend pos=north east,
]

\addplot[color=black, semithick,] table {
0	2.310492156
0.1	2.2608671086
0.2	2.2119918648
0.3	2.1638659189
0.4	2.116488429
0.5	2.0698582183
0.6	2.0239737776
0.7	1.9788332678
0.8	1.9344345233
0.9	1.8907750556
1	1.8478520579
1.1	1.8056624099
1.2	1.7642026835
1.3	1.7234691483
1.4	1.6834577784
1.5	1.6441642593
1.6	1.6055839947
1.7	1.5677121147
1.8	1.5305434838
1.9	1.4940727091
2	1.4582941492
2.1	1.4232019234
2.2	1.3887899205
2.3	1.3550518088
2.4	1.3219810458
2.5	1.2895708878
2.6	1.2578144001
2.7	1.2267044674
2.8	1.1962338035
2.9	1.1663949623
3	1.1371803474
3.1	1.108582223
3.2	1.0805927242
3.3	1.0532038666
3.4	1.0264075574
3.5	1.0001956048
3.6	0.9745597284
3.7	0.9494915688
3.8	0.9249826972
3.9	0.9010246255
4	0.8776088148
4.1	0.8547266851
4.2	0.8323696239
4.3	0.8105289951
4.4	0.7891961471
4.5	0.7683624214
4.6	0.74801916
4.7	0.7281577134
4.8	0.7087694477
4.9	0.689845752
5	0.6713780449
5.1	0.6533577809
5.2	0.6357764571
5.3	0.6186256185
5.4	0.6018968641
5.5	0.5855818519
5.6	0.5696723044
5.7	0.5541600127
5.8	0.5390368415
5.9	0.5242947332
6	0.5099257114
6.1	0.4959218853
6.2	0.4822754525
6.3	0.4689787022
6.4	0.4560240183
6.5	0.4434038818
6.6	0.4311108736
6.7	0.4191376759
6.8	0.407477075
6.9	0.3961219623
7	0.3850653363
7.1	0.3743003037
7.2	0.3638200807
7.3	0.3536179934
7.4	0.3436874795
7.5	0.3340220877
7.6	0.3246154794
7.7	0.3154614278
7.8	0.306553819
7.9	0.2978866512
8	0.2894540352
8.1	0.2812501936
8.2	0.2732694609
8.3	0.2655062826
8.4	0.2579552149
8.5	0.2506109238
8.6	0.2434681846
8.7	0.2365218806
8.8	0.2297670027
8.9	0.2231986478
9	0.2168120184
9.1	0.2106024209
9.2	0.2045652644
9.3	0.1986960601
9.4	0.1929904193
9.5	0.1874440523
9.6	0.1820527673
9.7	0.1768124686
9.8	0.1717191553
9.9	0.1667689202
10	0.1619579475
10.1	0.1572825124
10.2	0.1527389785
10.3	0.1483237972
10.4	0.1440335055
10.5	0.1398647249
10.6	0.1358141594
10.7	0.1318785946
10.8	0.1280548955
10.9	0.1243400053
11	0.1207309441
11.1	0.1172248067
11.2	0.1138187617
11.3	0.1105100496
11.4	0.1072959816
11.5	0.1041739377
11.6	0.1011413659
11.7	0.0981957799
11.8	0.0953347583
11.9	0.0925559429
12	0.0898570371
12.1	0.0872358052
12.2	0.08469007
12.3	0.0822177124
12.4	0.0798166695
12.5	0.0774849334
12.6	0.0752205499
12.7	0.0730216173
12.8	0.0708862853
12.9	0.0688127531
13	0.0667992691
13.1	0.064844129
13.2	0.062945675
13.3	0.0611022944
13.4	0.0593124188
13.5	0.0575745225
13.6	0.0558871222
13.7	0.0542487749
13.8	0.0526580778
13.9	0.0511136668
14	0.0496142156
14.1	0.0481584347
14.2	0.0467450705
14.3	0.0453729043
14.4	0.0440407514
14.5	0.0427474604
14.6	0.041491912
14.7	0.0402730184
14.8	0.0390897221
14.9	0.0379409959
15	0.0368258409
15.1	0.035743287
15.2	0.0346923911
15.3	0.0336722371
15.4	0.0326819345
15.5	0.0317206184
15.6	0.0307874484
15.7	0.0298816079
15.8	0.0290023035
15.9	0.0281487648
16	0.0273202428
};
\addlegendentry{$f_{\alpha=0.3}(x)$}

\addplot[
   color=black,
   %dash pattern={on 5pt off 2pt on 4pt off 1pt},
   dotted,
   thick,
]
table {
0	1.877
0.1	1.8568
0.2	1.8366
0.3	1.8164
0.4	1.7962
0.5	1.776
0.6	1.7558
0.7	1.7356
0.8	1.7154
0.9	1.6952
1	1.675
1.1	1.6548
1.2	1.6346
1.3	1.6144
1.4	1.5942
1.5	1.574
1.6	1.5538
1.7	1.5336
1.8	1.5134
1.9	1.4932
2	1.473
2.1	1.4528
2.2	1.4326
2.3	1.4124
2.4	1.3922
2.5	1.372
2.6	1.3518
2.7	1.3316
2.8	1.3114
2.9	1.2912
3	1.271
3.1	1.2508
3.2	1.2306
3.3	1.2104
3.4	1.1902
3.5	1.17
3.6	1.1498
3.7	1.1296
3.8	1.1094
3.9	1.0892
4	1.069
4.1	1.0488
4.2	1.0286
4.3	1.0084
4.4	0.9882
4.5	0.968
4.6	0.9478
4.7	0.9276
4.8	0.9074
4.9	0.8872
5	0.867
5.1	0.8468
5.2	0.8266
5.3	0.8064
5.4	0.7862
5.5	0.766
5.6	0.7458
5.7	0.7256
5.8	0.7054
5.9	0.6852
6	0.665
6.1	0.6448
6.2	0.6246
6.3	0.6044
6.4	0.5842
6.5	0.564
6.6	0.5438
6.7	0.5236
6.8	0.5034
6.9	0.4832
7	0.463
7.1	0.4428
7.2	0.4226
7.3	0.4024
7.4	0.3822
7.5	0.362
7.6	0.3418
7.7	0.3216
7.8	0.3014
7.9	0.2812
8	0.261
8.1	0.2408
8.2	0.2206
8.3	0.2004
8.4	0.1802
8.5	0.16
8.6	0.1398
8.7	0.1196
8.8	0.0994
8.9	0.0792
9	0.059
9.1	0.0388
9.2	0.0186
9.3	0
9.4	0
9.5	0
9.6	0
9.7	0
9.8	0
9.9	0
10	0
10.1	0
10.2	0
10.3	0
10.4	0
10.5	0
10.6	0
10.7	0
10.8	0
10.9	0
11	0
11.1	0
11.2	0
11.3	0
11.4	0
11.5	0
11.6	0
11.7	0
11.8	0
11.9	0
12	0
12.1	0
12.2	0
12.3	0
12.4	0
12.5	0
12.6	0
12.7	0
12.8	0
12.9	0
13	0
13.1	0
13.2	0
13.3	0
13.4	0
13.5	0
13.6	0
13.7	0
13.8	0
13.9	0
14	0
14.1	0
14.2	0
14.3	0
14.4	0
14.5	0
14.6	0
14.7	0
14.8	0
14.9	0
15	0
15.1	0
15.2	0
15.3	0
15.4	0
15.5	0
15.6	0
15.7	0
15.8	0
15.9	0
16	0

};
\addlegendentry{$f^{\text{lin}}_{\alpha=0.3}(x)$}

\addplot[Paired-1, dashed, thick] table {
0	1.5
5	1.5
5	0
16	0
};
\addlegendentry{$f^{*}_{\alpha=0.3}(x)$}

\end{axis}
\end{tikzpicture}
  \vspace{-1em}
  \caption{\label{fig:metricApprox} $f_{\alpha}(x)$ with $\alpha =0.3$, and its constant and linear approximations $f^*_\alpha(x)$ and $f^{\text{lin}}_{\alpha=0.3}(x)$, respectively.}
  \vspace{-1em}
\end{figure}
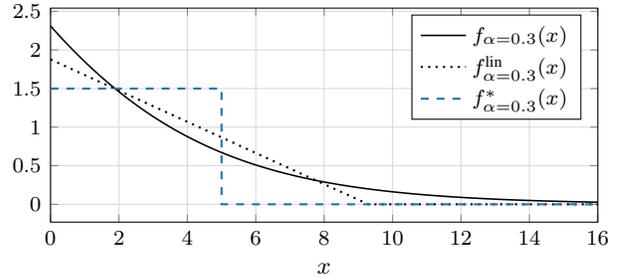

\begin{figure*}[t]
  \centering
  %!TEX root = ../FastDSCF.tex
\begin{tikzpicture}[spy using outlines=
    {circle, magnification=1.6, connect spies}]
    \begin{semilogyaxis}[
            footnotesize, width=2*\columnwidth, height=.70\columnwidth,    
            xmin=-7, xmax=8.25, xtick={-8,-7, ..., 9},
            ymin=5e-8,  ymax=2,
            xlabel=SNR (dB),%$\frac{E_s}{N_0} \text{~(dB)}$, %xlabel style={yshift=0.4em},
            ylabel=FER,  %ylabel style={yshift=-0.75em},
            grid=both, grid style={gray!30},
            tick align=outside, tickpos=left, 
            legend columns =1,
            legend pos=south west,
            legend style={font=\scriptsize},
            mark size = 3,
        ]

        \addplot[Paired-1, mark=o, semithick] table [x=SNR,y=DSCFA] {figures/data/r1-8_w1};
        \addplot[Paired-3, mark=triangle, semithick] table [x=SNR,y=DSCFL] {figures/data/r1-8_w1};
        \addplot[Paired-5, mark=x, semithick] table [x=SNR,y=DSCF ] {figures/data/r1-8_w1};
        \addplot[black   , dashed, semithick] table [x=SNR,y=SCO  ] {figures/data/r1-8_w1};

        \addplot[Paired-1, mark=o, semithick] table [x=SNR,y=DSCFA] {figures/data/r1-8_w2};
        \addplot[Paired-3, mark=triangle, semithick] table [x=SNR,y=DSCFL] {figures/data/r1-8_w2};
        \addplot[Paired-5, mark=x, semithick] table [x=SNR,y=DSCF ] {figures/data/r1-8_w2};
        \addplot[black   , dashed, semithick] table [x=SNR,y=SCO  ] {figures/data/r1-8_w2};

        \addplot[Paired-1, mark=o, semithick] table [x=SNR,y=DSCFA] {figures/data/r1-8_w3};
        \addplot[Paired-3, mark=triangle, semithick] table [x=SNR,y=DSCFL] {figures/data/r1-8_w3};
        \addplot[Paired-5, mark=x, semithick] table [x=SNR,y=DSCF ] {figures/data/r1-8_w3};
        \addplot[black   , dashed, semithick] table [x=SNR,y=SCO  ] {figures/data/r1-8_w3};

        \addplot[Paired-1, mark=o, semithick] table [x=SNR,y=DSCFA] {figures/data/r1-4_w1};
        \addplot[Paired-3, mark=triangle, semithick] table [x=SNR,y=DSCFL] {figures/data/r1-4_w1};
        \addplot[Paired-5, mark=x, semithick] table [x=SNR,y=DSCF ] {figures/data/r1-4_w1};
        \addplot[black   , dashed, semithick] table [x=SNR,y=SCO  ] {figures/data/r1-4_w1};

        \addplot[Paired-1, mark=o, semithick] table [x=SNR,y=DSCFA] {figures/data/r1-4_w2};
        \addplot[Paired-3, mark=triangle, semithick] table [x=SNR,y=DSCFL] {figures/data/r1-4_w2};
        \addplot[Paired-5, mark=x, semithick] table [x=SNR,y=DSCF ] {figures/data/r1-4_w2};
        \addplot[black   , dashed, semithick] table [x=SNR,y=SCO  ] {figures/data/r1-4_w2};

        \addplot[Paired-1, mark=o, semithick] table [x=SNR,y=DSCFA] {figures/data/r1-4_w3};
        \addplot[Paired-3, mark=triangle, semithick] table [x=SNR,y=DSCFL] {figures/data/r1-4_w3};
        \addplot[Paired-5, mark=x, semithick] table [x=SNR,y=DSCF ] {figures/data/r1-4_w3};
        \addplot[black   , dashed, semithick] table [x=SNR,y=SCO  ] {figures/data/r1-4_w3};

        \addplot[Paired-1, mark=o, semithick] table [x=SNR,y=DSCFA] {figures/data/r1-2_w1};
        \addplot[Paired-3, mark=triangle, semithick] table [x=SNR,y=DSCFL] {figures/data/r1-2_w1};
        \addplot[Paired-5, mark=x, semithick] table [x=SNR,y=DSCF ] {figures/data/r1-2_w1};
        \addplot[black   , dashed, semithick] table [x=SNR,y=SCO  ] {figures/data/r1-2_w1};

        \addplot[Paired-1, mark=o, semithick] table [x=SNR,y=DSCFA] {figures/data/r1-2_w2};
        \addplot[Paired-3, mark=triangle, semithick] table [x=SNR,y=DSCFL] {figures/data/r1-2_w2};
        \addplot[Paired-5, mark=x, semithick] table [x=SNR,y=DSCF ] {figures/data/r1-2_w2};
        \addplot[black   , dashed, semithick] table [x=SNR,y=SCO  ] {figures/data/r1-2_w2};

        \addplot[Paired-1, mark=o, semithick] table [x=SNR,y=DSCFA] {figures/data/r1-2_w3};
        \addplot[Paired-3, mark=triangle, semithick] table [x=SNR,y=DSCFL] {figures/data/r1-2_w3};
        \addplot[Paired-5, mark=x, semithick] table [x=SNR,y=DSCF ] {figures/data/r1-2_w3};
        \addplot[black   , dashed, semithick] table [x=SNR,y=SCO  ] {figures/data/r1-2_w3};

        \addplot[Paired-1, mark=o, semithick] table [x=SNR,y=DSCFA] {figures/data/r3-4_w1};
        \addplot[Paired-3, mark=triangle, semithick] table [x=SNR,y=DSCFL] {figures/data/r3-4_w1};
        \addplot[Paired-5, mark=x, semithick] table [x=SNR,y=DSCF ] {figures/data/r3-4_w1};
        \addplot[black   , dashed, semithick] table [x=SNR,y=SCO  ] {figures/data/r3-4_w1};

        \addplot[Paired-1, mark=o, semithick] table [x=SNR,y=DSCFA] {figures/data/r3-4_w2};
        \addplot[Paired-3, mark=triangle, semithick] table [x=SNR,y=DSCFL] {figures/data/r3-4_w2};
        \addplot[Paired-5, mark=x, semithick] table [x=SNR,y=DSCF ] {figures/data/r3-4_w2};
        \addplot[black   , dashed, semithick] table [x=SNR,y=SCO  ] {figures/data/r3-4_w2};

        \addplot[Paired-1, mark=o, semithick] table [x=SNR,y=DSCFA] {figures/data/r3-4_w3};
        \addplot[Paired-3, mark=triangle, semithick] table [x=SNR,y=DSCFL] {figures/data/r3-4_w3};
        \addplot[Paired-5, mark=x, semithick] table [x=SNR,y=DSCF ] {figures/data/r3-4_w3};
        \addplot[black   , dashed, semithick] table [x=SNR,y=SCO  ] {figures/data/r3-4_w3};

        \addplot[Paired-1, mark=o, semithick] table [x=SNR,y=DSCFA] {figures/data/r7-8_w1};
        \addplot[Paired-3, mark=triangle, semithick] table [x=SNR,y=DSCFL] {figures/data/r7-8_w1};
        \addplot[Paired-5, mark=x, semithick] table [x=SNR,y=DSCF ] {figures/data/r7-8_w1};
        \addplot[black   , dashed, semithick] table [x=SNR,y=SCO  ] {figures/data/r7-8_w1};

        \addplot[Paired-1, mark=o, semithick] table [x=SNR,y=DSCFA] {figures/data/r7-8_w2};
        \addplot[Paired-3, mark=triangle, semithick] table [x=SNR,y=DSCFL] {figures/data/r7-8_w2};
        \addplot[Paired-5, mark=x, semithick] table [x=SNR,y=DSCF ] {figures/data/r7-8_w2};
        \addplot[black   , dashed, semithick] table [x=SNR,y=SCO  ] {figures/data/r7-8_w2};

        \addplot[Paired-1, mark=o, semithick] table [x=SNR,y=DSCFA] {figures/data/r7-8_w3};
        \addplot[Paired-3, mark=triangle, semithick] table [x=SNR,y=DSCFL] {figures/data/r7-8_w3};
        \addplot[Paired-5, mark=x, semithick] table [x=SNR,y=DSCF ] {figures/data/r7-8_w3};
        \addplot[black   , dashed, semithick] table [x=SNR,y=SCO  ] {figures/data/r7-8_w3};

        \legend{DSCF using $f_\alpha^*$, DSCF using $f^{\text{lin}}_\alpha$, DSCF using $f_\alpha$, SC-Oracle}

        \node [color=black] at (axis cs:-4.50,3e-01) {\footnotesize $R=\frac{1}{8}$};
        \node [color=black] at (axis cs:-1.50,3e-01) {\footnotesize $R=\frac{1}{4}$};
        \node [color=black] at (axis cs:2.25,3e-01) {\footnotesize $R=\frac{1}{2}$};
        \node [color=black] at (axis cs:3.8,5e-02) {\footnotesize $R=\frac{3}{4}$};
        \node [color=black] at (axis cs:6.75,3e-01) {\footnotesize $R=\frac{7}{8}$};

        \draw[->] (axis cs:-1.9,1.5e-7) -- (axis cs:-3.0,7e-8) node [right=10pt] {\footnotesize $\omega$};
        \draw[->] (axis cs:0.7,4e-7) -- (axis cs:-0.2,1e-7) node [right=10pt] {\footnotesize $\omega$};
        \draw[->] (axis cs:3.8,4e-7) -- (axis cs:2.9,1.5e-7) node [right=10pt] {\footnotesize $\omega$};
        \draw[->] (axis cs:6.9,2e-7) -- (axis cs:5.9,7e-8) node [right=10pt] {\footnotesize $\omega$};
        \draw[->] (axis cs:8.2,2e-6) -- (axis cs:7.5,3e-7) node [right=5pt] {\footnotesize $\omega$};
         % \draw (axis cs:0,0) ellipse [x radius=1cm, y radius=1cm];

        \coordinate (spypoint1) at (axis cs:-3.75,1.3e-4);
        \coordinate (magnifyglass1) at (axis cs:-5.50,1.3e-4);

        \coordinate (spypoint2) at (axis cs:-0.80,1.3e-4);
        \coordinate (magnifyglass2) at (axis cs:-2.15,1.3e-4);

        \coordinate (spypoint3) at (axis cs:2.60,1.3e-4);
        \coordinate (magnifyglass3) at (axis cs:1.00,1.3e-4);

        \coordinate (spypoint4) at (axis cs:5.60,1.0e-4);
        \coordinate (magnifyglass4) at (axis cs:4.00,1e-4);

        \coordinate (spypoint5) at (axis cs:7.05,1.0e-4);
        \coordinate (magnifyglass5) at (axis cs:5.90,5e-3);

    \end{semilogyaxis}
    \spy [Paired-11, height=1.5cm, width=1.5cm] on (spypoint1)
   in node[fill=white] at (magnifyglass1);
    \spy [Paired-11, height=1.5cm, width=1.5cm] on (spypoint2)
   in node[fill=white] at (magnifyglass2);
    \spy [Paired-11, height=1.5cm, width=1.5cm] on (spypoint3)
   in node[fill=white] at (magnifyglass3);
    \spy [Paired-11, height=1.5cm, width=1.5cm] on (spypoint4)
   in node[fill=white] at (magnifyglass4);
    \spy [Paired-11, height=1.0cm, width=1.0cm] on (spypoint5)
   in node[fill=white] at (magnifyglass5);

\end{tikzpicture} 
  \caption{\label{fig:metricCompare_FER} FER performance comparison of DSCF decoding with and without approximations, using $\alpha =0.3$. Polar codes with $N=1024$ and $R \in \{\frac{1}{8}, \frac{1}{4}, \frac{1}{2}, \frac{3}{4}, \frac{7}{8} \}$, $C=16$, $\omega \in \{1,2,3\}$ with $T_{\text{max}} \in \{10,40,200\}$.}
\end{figure*}
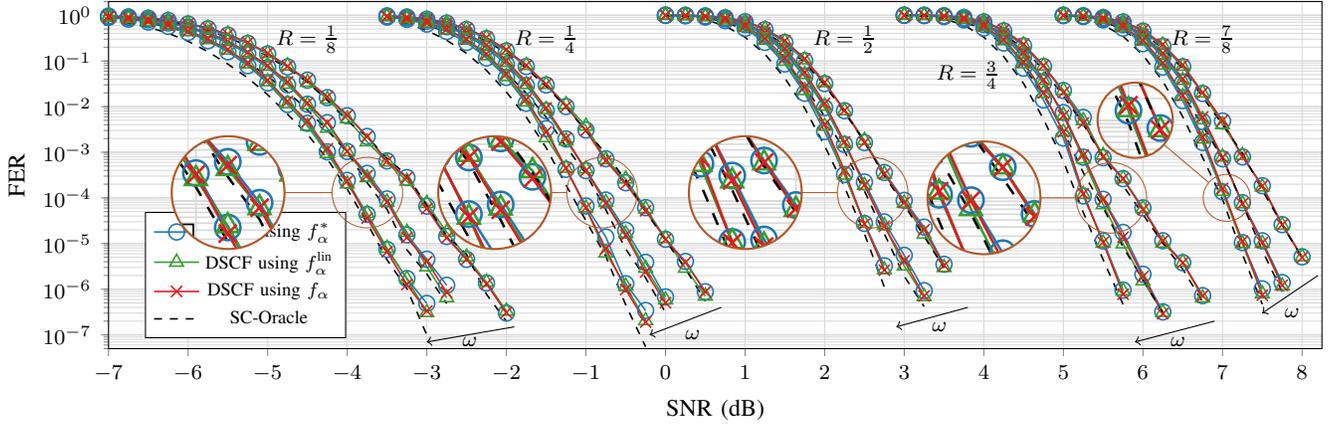

Following the Monte-Carlo optimizations from \cite{SCFlip17-jour}, $\alpha = 0.3$ is used throughout this paper. Interestingly, it was shown that a similar expression, $f(x) = \log\big{(}1+\exp(-x)\big{)}$, used in the soft-input soft-output decoding algorithm of turbo codes, can be approximated in different ways without adversely affecting the decoding performance \cite{Gross_logMAP_1998,Cheng_logMAP_2000}. Inspired from the \textit{constant log-MAP} approximation in \cite{Gross_logMAP_1998}, we use a similar approximation to simplify $f_{\alpha}(x)$ as:
\begin{equation}\label{eqn:fxconst}
f^{*}_{\alpha=0.3}(x) = \left\{
  \begin{array}{@{}ll@{}}
    \frac{3}{2}, & \text{if}~ |x| \leq 5 \\
    0, & \text{otherwise.}
  \end{array}\right.
\end{equation}
The values of $\frac{3}{2}$ and $5$ in (\ref{eqn:fxconst}) are selected to ensure both an easy future hardware implementation and a reduced fitting error. For illustration purpose, Fig.~\ref{fig:metricApprox} plots the original function $f_\alpha$ and its proposed approximation $f^*_\alpha$, with $\alpha =0.3$. Note that in \cite{Zhou_improvedFastSCF}, a \textit{linear approximation} to $f_{\alpha}(x)$ was used following \cite{Cheng_logMAP_2000} to reduce its complexity. An illustration for the linear approximation is also depicted in Fig. \ref{fig:metricApprox}, labeled as $f^{\text{lin}}_{\alpha=0.3}(x)$. However, our approach to approximate $f_{\alpha}(x)$ is simpler as it only involves a constant value. Fig.~\ref{fig:metricCompare_FER} compares the FER performance of DSCF decoding with the constant and linear approximations against its original approach from \cite{SCFlip17-jour}, using length-1024 polar codes with different rates. The polar codes are constructed using the 5G reliability sequence \cite{38.212}. The length-16 CRC defined in \cite{38.212} is serially concatenated with the polar code. A BPSK modulation and an AWGN channel are considered. Note that the same settings are used for all the following Monte-Carlo simulations. Three error orders $\omega \in \{1,2,3\}$ are targeted, corresponding to $T_{\text{max}} \in \{10,40,200\}$, respectively. The decoding performance with SC-Oracle for each $\omega$ value is also shown. Observe that  in the considered cases, the approximated DSCF curves achieve similar decoding performance as the original approach but without transcendental computations. As the constant approximation is more favorable for reduced complexity, we choose to replace $f_{\alpha}(x)$ with $f^{*}_{\alpha}(x)$.

\section{Fast-DSCF Decoding}\label{sec:fastdscf}

\subsection{Achievable Performance by the Fast-SCF-based Decoders}\label{subsec:fastsco}

We introduce the concept of \textit{Fast-SCO} decoder to depict the ideal limit on the achievable performance by SCF-based decoders when special nodes that enclose more than one non-frozen index are incorporated. The modified SCO decoder works exactly as a Fast-SSC decoder, except that it is able to identify and correct up to a certain number of channel-induced errors at the top-level of the special nodes. The introduction of Fast-SCO is essential towards the performance evaluation of the proposed \textit{Fast-DSCF} decoder.

The two special nodes of our interest, which involves more than a single non-frozen index, are Rate-1 and SPC nodes. Note that the FER value of Fast-SCO is the same as SCO when only Rate-0 and Rep nodes are involved. For Rate-1 nodes, up to $\omega$ channel-induced errors are found and flipped. On the other hand, an SPC node involves a single frozen index, and its parity must be kept even at all times. As such, an even number of bit-flips have to take place at SPC nodes in order to keep their even parity constraint. Therefore, two simultaneous bit-flips have to be performed in SPC nodes, for each error order. 

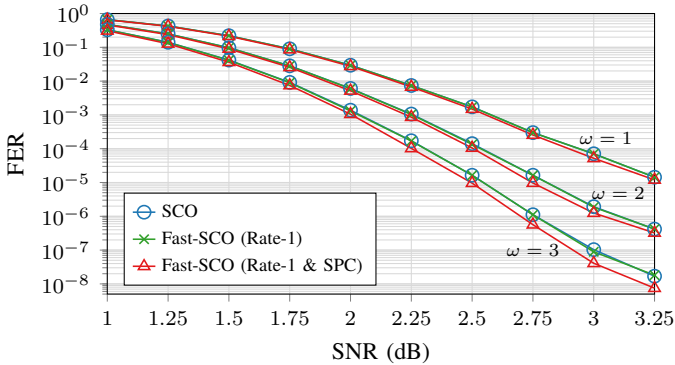
\begin{figure}[t]
  \centering
  %!TEX root = ../FastDSCF.tex

\begin{tikzpicture}[spy using outlines=
	{circle, magnification=2.0, connect spies}]

    \begin{semilogyaxis}[
            footnotesize, width=\columnwidth, height=.60\columnwidth,    
            xmin=1, xmax=3.25, xtick={1, 1.25,...,3.25},
            ymin=5e-9,  
            ymax=1,
            xlabel=SNR (dB),%$\frac{E_s}{N_0} \text{~(dB)}$, %xlabel style={yshift=0.4em},
            ylabel=FER,  %ylabel style={yshift=-0.75em},
            grid=both, grid style={gray!30},
            tick align=outside, tickpos=left, 
            mark size = 2.5,
            legend cell align={left},
            legend columns=1,
            legend pos=south west,
            legend image code/.code={
                \draw[mark repeat=2,mark phase=2]
                plot coordinates {
                (0cm,0cm)
                (0.15cm,0cm)        %% default is (0.3cm,0cm)
                (0.3cm,0cm)         %% default is (0.6cm,0cm)
                };%
                },
            legend style={font=\small},
        ]

%\addlegendimage{empty legend}
%\addlegendentry{SCO}

\addplot[
    color=Paired-1,
    mark=o,
    semithick
]
table {
1.00 6.59430e-01
1.25 4.27870e-01
1.50 2.20750e-01
1.75 9.03500e-02
2.00 2.95000e-02
2.25 7.42000e-03
2.50 1.73000e-03
2.75 2.99287e-04
3.00 7.08938e-05
3.25 1.43373e-05
3.50 2.92422e-06
3.75 6.66162e-07
4.00 1.30563e-07
};
%\addlegendentry{$w=1$}

%\addlegendimage{empty legend}
%\addlegendentry{Fast-SCO (Rate-1)}

%w=1,t=10
\addplot[
    color=Paired-3,
    mark=x,
    semithick
]
table {
1.00e+00 6.58700e-01
1.25e+00 4.26950e-01
1.50e+00 2.19980e-01
1.75e+00 8.98900e-02
2.00e+00 2.92900e-02
2.25e+00 7.38000e-03
2.50e+00 1.72000e-03
2.75e+00 2.99287e-04
3.00e+00 7.05283e-05
3.25e+00 1.43373e-05
};
%\addlegendentry{$w=1$}

%\addlegendimage{empty legend}
%\addlegendentry{Fast-SCO (Rate-1 \& SPC)}

\addplot[
    color=Paired-5,
    mark=triangle,
    semithick
]
table {
1.00e+00 6.51470e-01
1.25e+00 4.19330e-01
1.50e+00 2.13600e-01
1.75e+00 8.60200e-02
2.00e+00 2.75400e-02
2.25e+00 6.74000e-03
2.50e+00 1.49000e-03
2.75e+00 2.54723e-04
3.00e+00 5.20794e-05
3.25e+00 1.19762e-05
};
%\addlegendentry{$w=1$}

\addplot[
    color=Paired-1,
    mark=o,
    semithick
]
table {
1.00 4.76790e-01
1.25 2.50240e-01
1.50 9.71400e-02
1.75 2.83400e-02
2.00 6.06000e-03
2.25 1.06000e-03
2.50 1.41300e-04
2.75 1.64007e-05
3.00 1.92590e-06
3.25 4.18208e-07
};
%\addlegendentry{$w=2$}

%w=2,t=40
\addplot[
    color=Paired-3,
    mark=x,
    semithick,
]
table {
1.00e+00 4.75160e-01
1.25e+00 2.48700e-01
1.50e+00 9.63500e-02
1.75e+00 2.81400e-02
2.00e+00 5.97000e-03
2.25e+00 1.03000e-03
2.50e+00 1.34443e-04
2.75e+00 1.63173e-05
3.00e+00 1.85681e-06
3.25e+00 4.18208e-07
};
%\addlegendentry{$w=2$}

\addplot[
    color=Paired-5,
    mark=triangle,
    semithick
]
table {
1.00e+00 4.60470e-01
1.25e+00 2.36820e-01
1.50e+00 8.98300e-02
1.75e+00 2.56400e-02
2.00e+00 5.17000e-03
2.25e+00 8.60000e-04
2.50e+00 1.05151e-04
2.75e+00 9.66288e-06
3.00e+00 1.24937e-06
3.25e+00 3.22478e-07
};
%\addlegendentry{$w=2$}

\addplot[
    color=Paired-1,
    mark=o,
    semithick
]
table {
1.00 3.30310e-01
1.25 1.40990e-01
1.50 4.19600e-02
1.75 9.17000e-03
2.00 1.38000e-03
2.25 1.72022e-04
2.50 1.63814e-05
2.75 1.11383e-06
3.00 1.03196e-07
3.25 1.69226e-08
};
%\addlegendentry{$w=3$}
\addlegendentry{\scriptsize SCO}

%w=3,t=200
\addplot[
    color=Paired-3,
    mark=x,
    semithick
]
table {
1.00e+00 3.28000e-01
1.25e+00 1.39490e-01
1.50e+00 4.13000e-02
1.75e+00 8.98000e-03
2.00e+00 1.31000e-03
2.25e+00 1.68725e-04
2.50e+00 1.63814e-05
2.75e+00 1.10778e-06
3.00e+00 8.91652e-08
3.25e+00 1.81197e-08
};
%\addlegendentry{$w=3$}
\addlegendentry{\scriptsize Fast-SCO (Rate-1)}

\addplot[
    color=Paired-5,
    mark=triangle,
    semithick
]
table {
1.00e+00 3.08810e-01
1.25e+00 1.27510e-01
1.50e+00 3.63200e-02
1.75e+00 7.45000e-03
2.00e+00 1.05000e-03
2.25e+00 1.02257e-04
2.50e+00 9.58477e-06
2.75e+00 5.65421e-07
3.00e+00 4.08389e-08
3.25e+00 7.49206e-09
};
%\addlegendentry{$w=3$}
\addlegendentry{\scriptsize Fast-SCO (Rate-1 \& SPC)}

\node [color=black] at (axis cs:3.05,2e-04) {\scriptsize $\omega=1$};
\node [color=black] at (axis cs:3.10,5e-06) {\scriptsize $\omega=2$};
\node [color=black] at (axis cs:2.75,1e-07) {\scriptsize $\omega=3$};

\end{semilogyaxis}
\end{tikzpicture}
  \vspace{-2em}
  \caption{\label{fig:FER_SCO} FER performance limits depicted by SCO and Fast-SCO when Rate-1 and SPC nodes are involved, using $PC(1024,512)$ and $C=16$.}%0x1021
\end{figure}

Fig.~\ref{fig:FER_SCO} compares the FER performance limits depicted by SCO against Fast-SCO, when Rate-1 or Rate-1 and SPC nodes are involved for $\omega \in \{1,2,3\}$. It can be noticed that the performance limit of Fast-SCO with only Rate-1 nodes is equivalent to that of SCO, and the Fast-SCO adaptation involving SPC nodes has an improved performance limit compared to others. This is because a portion of corrected errors at the top of the SPC node correspond to multiple channel-induced error-corrections at the leaf node level. Hence, a better performance can be ideally achieved when SPC nodes are used in SCF-based decoders.

\subsection{Incorporation of Special Nodes into DSCF}\label{subsec:fastnodes}

The main task in incorporating the special nodes into DSCF decoding is to conceive metric computations (\ref{eqn:dscfmetricM})-(\ref{eqn:fxconst}) using the available information at the top-level of the special nodes. In the following, we present how to perform the metric computations for each special node. Note that the following approaches can be applied to DSCF decoding with or without the approximation presented in Section~\ref{sec:metricapprox}.

Let us split the metric calculation of DSCF into two parts, such that:
\begin{equation}\label{eqn:dscfmetricMnew}
M_\alpha(\mathcal{E}_{\omega}) = \underbrace{|L^{0}[\mathcal{E}_{\omega-1}]_{i_\omega}|}_{M'_{\alpha}(\mathcal{E}_{\omega})} + \underbrace{\Big{(}\sum_{j \in \mathcal{E}_{\omega-1}} | L^{0}[\mathcal{E}_{\omega-1}]_j | + S_{\alpha}(\mathcal{E}_{\omega}) \Big{)}}_{M''_{\alpha}(\mathcal{E}_{\omega})}.
\end{equation}
Observe that $M'_{\alpha}(\mathcal{E}_{\omega})$ takes a value only if the index $i_\omega$ is a candidate for bit-flipping during a future decoding iteration. $M'_{\alpha}(\mathcal{E}_{\omega})$ contains the \textit{instantaneous value} at the index $i_\omega$ and used for the metric computation of the index $i_\omega$ \textit{only}. On the other hand, $M''_{\alpha}(\mathcal{E}_{\omega})$ is the \textit{accumulative part} that is used for the next possible set of bit-flips throughout the decoding attempt. $M''_{\alpha}(\mathcal{E}_{\omega})$ is set to $0$ at the beginning of any extra decoding attempt and accumulated for each non-frozen leaf index $j$ as follows:
\begin{align}\label{eqn:dscfmetricMnew2update}
M''_{\alpha}(\mathcal{E}_{\omega})_j =&~  M''_{\alpha}(\mathcal{E}_{\omega})_{j-1}\notag\\
                                     &+ | L^{0}[\mathcal{E}_{\omega-1}]_j | \text{~if~}  j \in \mathcal{E}_{\omega-1}\notag\\
                                     &+ f_{\alpha}(| L^{0}[\mathcal{E}_{\omega-1}]_j|) \text{~if~} j \in \mathcal{A}.
\end{align}

We now provide a way to compute $M'_{\alpha}(\mathcal{E}_{\omega})$ and $M''_{\alpha}(\mathcal{E}_{\omega})$ when special nodes are encountered during DSCF decoding. First, we redefine the set $\mathcal{E}_{\omega}$ to hold the information of the special nodes as well as the flipping indices. Let us define the notation $\{j,\boldsymbol{i} \}$ as the \textit{coordinate} in a polar code tree where $j$ denotes the special node index and $\boldsymbol{i}$ denotes a set of top-node indices that belongs to $j$. Accordingly, let $\mathcal{E}_{\omega} = \{ \{j_1,\boldsymbol{i_1} \}, \dots,\{j_\omega,\boldsymbol{i_\omega} \}  \}$ denote the set of flipping coordinates at special nodes $\{j_1,\dots,j_\omega\}$ ($j_1 \leq \cdots \leq j_\omega$). In this context, a flipping coordinate $\{j,\boldsymbol{i}\}$ becomes a \textit{subset} of the set $\mathcal{E}_{\omega}$. Depending on the type of the special node, the cardinality of $\boldsymbol{i}$ can vary. As we will see, the instantaneous component of the metric computation $M'_{\alpha}(\mathcal{E}_{\omega})$ depends on the subset $\{j,\boldsymbol{i}\}$, whereas the accumulative component $M''_{\alpha}(\mathcal{E}_{\omega})$ is updated at once at each special node $j$.

The decoding of Rate-0 nodes for DSCF decoding is the same as \cite{SSC2011} since their LLR magnitudes are not evaluated towards metric updates; this has been addressed previously in \cite{FastSCFlip_WCNC18} for SCF-based algorithms.

Repetition (Rep) nodes contain a single non-frozen instance located at the rightmost index at the leaf node. From the top-level perspective, the information of the non-frozen index is distributed amongst all indices of the top node. In other words, the LLR in the last leaf node obtained through SC decoding is equal to the sum of all LLRs in the root node; and there is only one possible flipping-event. Therefore, for a Rep node $j$ of size $N_v$, at decoding tree stage $S$, we can write 
\begin{equation}\label{eqn:dscfmetricMnew1Rep}
M'_{\alpha}(\mathcal{E}_{\omega})_{\{j,\varnothing\}}  =  \bigg{|}\sum_{i \in N_v} L^{S}[\mathcal{E}_{\omega-1}]_i\bigg{|}, %\text{~, and}
\end{equation}
and the update for $M''_{\alpha}(\mathcal{E}_{\omega})$ at a Rep node can be expressed as
\begin{align}\label{eqn:dscfmetricMnew2updateRep}
M''_{\alpha}(\mathcal{E}_{\omega}) \pluseq &~ f_{\alpha}\Bigg{(}\bigg{|}\sum_{i \in N_v} L^{S}[\mathcal{E}_{\omega-1}]_i\bigg{|}\Bigg{)}\notag\\
                                     &+ \bigg{|}\sum_{i \in N_v} L^{S}[\mathcal{E}_{\omega-1}]_i\bigg{|} \text{~if~}  \{j,\varnothing\} \subset \mathcal{E}_{\omega-1}.%\notag\\
                                     % &
\end{align}
If a flipping event for a Rep node is selected during an extra decoding attempt, all the $N_v$ partial sums at level $S$ have to be flipped. Accordingly, there is no index information required for Rep nodes ($\boldsymbol{i} = \varnothing$ in $\{j,\boldsymbol{i}\}$). Note that the proposed metric calculation and update for Rep nodes are exact to the baseline DSCF decoding.

By definition, Rate-1 nodes do not involve any frozen bits. Thus, they correspond to an uncoded sequence and all the indices at the top of the node are evaluated for bit-flipping. We therefore use the top-level LLRs directly in metric calculations for the prospective flipping indices. 
For a Rate-1 node $j$ of size $N_v$, at decoding tree stage $S$, $M'_{\alpha}(\mathcal{E}_{\omega})$ is calculated for each index $i$ ($0 \leq i < N_v$) within the node, such that
\begin{equation}\label{eqn:dscfmetricMnew1Rate1}
M'_{\alpha}(\mathcal{E}_{\omega})_{\{j,i\}}  =  |L^{S}[\mathcal{E}_{\omega-1}]_{i}| \text{~.}
\end{equation}
Note that, the LLR values at the top of a Rate-1 node do not have any dependencies on one another. Therefore, $M''_{\alpha}(\mathcal{E}_{\omega})$ is updated \textit{at once} for the entire Rate-1 node, such that
\begin{align}\label{eqn:dscfmetricMnew2updateRate1}
M''_{\alpha}(\mathcal{E}_{\omega})  \pluseq&~ 
 \sum_{i \in N_v}    f_{\alpha}(| L^{S}[\mathcal{E}_{\omega-1}]_{i}|) \notag\\ & + \mkern-18mu  \sum_{\substack{ i \in N_v \\ \{j,i\} \subset \mathcal{E}_{\omega-1}}}  \mkern-18mu  | L^{S}[\mathcal{E}_{\omega-1}]_{i}| \text{~.}
\end{align}
If the flipping subset $\{j,i\}$ corresponds to a Rate-1 node at an additional decoding attempt, its top-node index $i$ is flipped. 

\begin{figure}[t]
  \centering
  \input{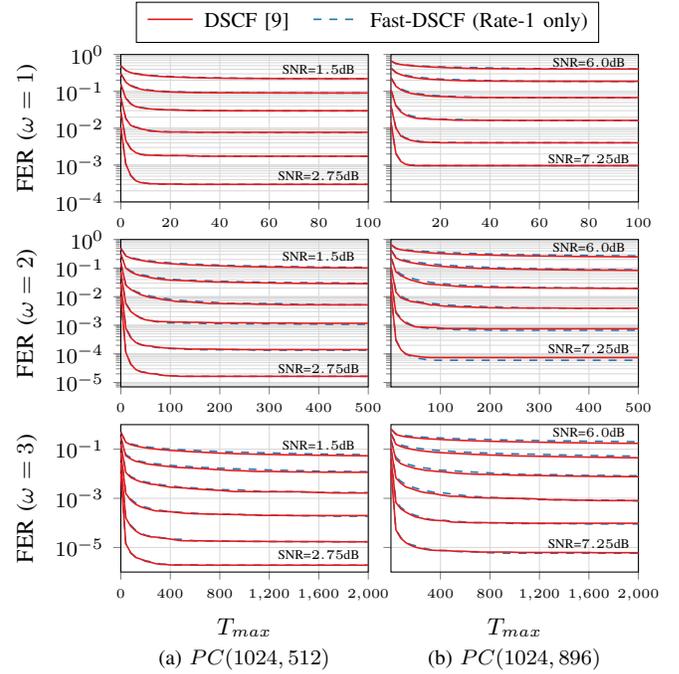}
  \vspace{-1em}
  \caption{\label{fig:TMAXvsFER_Rate1}FER performance of DSCF decoding with and without using Rate-1 nodes, with respect to a wide range of $T_{\text{max}}$ values. $C=16$, $\omega \in \{1,2,3\}$. $f_{\alpha=0.3}(x)$ is not approximated. Simulated SNR values are between $1.5$ dB and $2.75$ dB for $PC(1024,512)$, and between $6.0$ dB and $7.25$ dB for $PC(1024,896)$, with $0.25$ dB steps each.}
\end{figure}

The proposed metric for Rate-1 nodes is not exact to the baseline DSCF decoding. In return, Fig.~\ref{fig:TMAXvsFER_Rate1} depicts how the FER proceeds with $T_{\text{max}}$, using two polar codes $PC(1024,512)$ and $PC(1024,896)$, evaluated at error orders $\omega \in \{1,2,3\}$ and simulated at six different SNR points each. The non-frozen set $\mathcal{A}$ associated with these polar codes exhibit $95\%$ and $99\%$ of the indices that fall under Rate-1 nodes. It can be seen that the error-correction performance with Rate-1 nodes is similar to that of the original DSCF algorithm. In return, the decoding tree does not need to be traversed at Rate-1 nodes.

As mentioned in Section~\ref{subsec:fastsco}, an even number of bit-flips have to take place at SPC nodes in order to keep their even parity constraint. Accordingly, for each attempted error order, two indices are considered, which leads to $\binom{N_v}{2}$ combinations for an SPC node of size $N_v$. Therefore, for an SPC node $j$ of size $N_v$, at decoding tree stage $S$, $M'_{\alpha}(\mathcal{E}_{\omega})$ is calculated for each $\{j,\boldsymbol{i}\}$ ($\boldsymbol{i} = \{i_1,i_2\}, 0 \leq i_1 < N_v$, $0 \leq i_2 < N_v$, $i_1 \neq i_2$), such that
\begin{equation}\label{eqn:dscfmetricMnew1SPC}
M'_{\alpha}(\mathcal{E}_{\omega})_{\{j,\{i_1,i_2\}\}}  = \sum_{i \in \{i_1,i_2\}} \Big{(} |L^{S}[\mathcal{E}_{\omega-1}]_{i}| - \gamma |L^{S}[\mathcal{E}_{\omega-1}]_{i_{\text{min}}}| \Big{)} \text{~,}
\end{equation}
where $i_{\text{min}}$ denotes the top-node index with the minimum LLR magnitude and $\gamma$ is the initial parity.

As noted in (\ref{eqn:dscfmetricS}), only the non-frozen index LLR magnitudes are used towards the calculation of $S_{\alpha}(\mathcal{E}_{\omega})$. Thus, the index $i_{\text{min}}$ is excluded in the calculation of $S_{\alpha}(\mathcal{E}_{\omega})$ in SPC nodes. Instead, its LLR magnitude is applied as an offset to all other indices, such that
\begin{align}\label{eqn:dscfmetricMnew2updateSPC}
M''_{\alpha}(\mathcal{E}_{\omega})  \pluseq&~ 
 \mkern-18mu  \sum_{{ i \in N_v \setminus i_{\text{min}} }} \mkern-18mu  f_{\alpha}\big{(}| L^{S}[\mathcal{E}_{\omega-1}]_{i}| + (1-2\gamma) | L^{S}[\mathcal{E}_{\omega-1}]_{i_{\text{min}}}| \big{)} \notag\\ & + \mkern-18mu  \sum_{\substack{ i \in \{i_1,i_2\}  \\ \{j,\{i_1,i_2\}\} \subset \mathcal{E}_{\omega-1}}}  \mkern-18mu  \Big{(} |L^{S}[\mathcal{E}_{\omega-1}]_{i}| - \gamma |L^{S}[\mathcal{E}_{\omega-1}]_{i_{\text{min}}}| \Big{)} \text{~.}
\end{align}

If the flipping subset $\{j,\boldsymbol{i}\}$ correspond to an SPC node at an additional decoding attempt, then the two top node indices in $\boldsymbol{i}$ are flipped.

\begin{figure}[t]
  \centering
  \input{figures/TMAXvsFER_SPC_group.tikz}
  \vspace{-1em}
  \caption{\label{fig:TMAXvsFER_SPC}FER trend line of DSCF decoding with and without using SPC nodes, with respect to a wide range of $T_{\text{max}}$ values. $C=16$, $\omega \in \{1,2,3\}$. $f_{\alpha=0.3}(x)$ is not approximated. Simulated SNR values are between $1.5$ B and $2.75$ dB for $PC(1024,512)$, and between $6.0$ dB and $7.25$ dB for $PC(1024,896)$, with $0.25$ dB steps each.}
\end{figure}

Fig.~\ref{fig:TMAXvsFER_SPC} compares the FER performance of our approach with SPC nodes (\ref{eqn:dscfmetricMnew1SPC})-(\ref{eqn:dscfmetricMnew2updateSPC}) against the baseline DSCF algorithm for $\omega \in \{1,2,3\}$. As in the Rate-1 study, $PC(1024,512)$ and $PC(1024,896)$ are used: their associated non-frozen set exhibit $68\%$ and $65\%$ of the indices that fall under SPC nodes, respectively. For both polar codes, it is observed that their FER performance are able to reach lower FER values when SPC nodes are involved as outlined by the Fast-SCO performance limits in Fig.~\ref{fig:FER_SCO}. On the other hand, it is also observed that with increasing $\omega$, the performance with SPC nodes are slightly degraded at low $T_{\text{max}}$ values and better performance is achieved at relatively higher SNR and $T_{\text{max}}$ values.

\section{Reducing the Bit-Flipping Search Span in Fast-DSCF Decoding}\label{sec:reducedsearch}

In this Section, we attempt to reduce the search span for bit-flipping within Rate-1 and SPC nodes without introducing significant degradation in error-correction performance using a theoretical framework. This study is divided into three parts. First, we derive the theoretical FER bounds for Rate-1 nodes for any error order via density evolution using Gaussian approximation\cite{Trifonov_polarcodedesign2012}. Then, we attempt to approach to the derived bounds using a reduced number of elements within the Rate-1 nodes. Finally, we extend our study towards SPC nodes.

Note that these derived theoretical computations are not exclusive to the DSCF algorithm, and can be used for other algorithms and purposes. This study is carried out using BPSK modulation, it can be extended to higher-order modulation scenarios.

\subsection{Reducing the Search Span in Rate-1 Nodes}\label{sec:reducedsearch:rate1}

Assuming an all-zero codeword with BPSK signaling, the LLRs can be expressed as Gaussian random variables, such that

\begin{equation}
L^{0}_{i} \sim\ \mathcal{N}(\mu_i,\sigma_i^2), i \in [0;N)\text{.}
\end{equation}

As the variance ($\sigma^2$) and the mean ($\mu$) of the random variable model are coupled ($\sigma^2 = 2\mu$), it is sufficient to track the mean for each channel \cite{Trifonov_polarcodedesign2012}. 
Then, the probability of error ($\pi_i$) associated with each leaf node index can be approximated by the \textit{Q-function} $Q(x)$, which is referred to as the tail probability of a Gaussian distribution. The Q-function can be described using the complementary error function \cite{proakis}:
\begin{equation}\label{eqn:pi}
\pi_i \approx Q\Big{(}\frac{\mu_i}{\sigma_i}\Big{)} = Q\Big{(}\sqrt{\frac{\mu_i}{2}}\Big{)} =  \frac{1}{2} \erfc \Big(\frac{\sqrt{\mu_i}}{2}\Big)\text{.}
\end{equation}

Recall that $\text{Pr}(\mathcal{E}_{\omega})$ is the probability of SC decoding being successful after flipping the indices in $\mathcal{E}_{\omega}$ (\ref{eqn:prob_dscf}). Here, let us define a slightly different probability, $\text{Pr}(e_{\omega})$, as SC decoding being successful \textit{after flipping $\omega$ indices}. Differently than $\text{Pr}(\mathcal{E}_{\omega})$, $\text{Pr}(e_{\omega})$ does not involve a specific set of indices but only concerns the number of flipped indices. In this sense, when $\omega=0$, then $\text{Pr}(e_{0})$ is the probability of SC decoding being successful, and can be expressed in terms of $\pi_i$ as 
\begin{equation}\label{eqn:FER_SC_complement}
\text{Pr}(e_{0}) = \prod_{i \in \mathcal{A}}(1-\pi_i)\text{,}
\end{equation}
Accordingly, the FER of SC can be easily computed by taking the complement of $\text{Pr}(e_{0})$ \cite{wu2014construction}, such that

\begin{equation}\label{eqn:FERtheory}
\text{FER}_{\text{SC}} = 1- \text{Pr}(e_{0})=  1 - \Big[ \prod_{i \in \mathcal{A}}(1-\pi_i) \Big] \text{.}
\end{equation}

It is worth to mention that, the theoretical FER of SC when an error (or more) is corrected ($\omega>0$) can be derived similarly (\textit{e.g.} $1- \text{Pr}(e_{0}) - \text{Pr}(e_{1})$). However, the derived performance does not correlate well with the simulated performance. This is because of the propagated errors that occur at the leaf nodes. Since no systematic model is derived to explain the behavior of the propagated errors, they are assumed unpredictable. The correlation between channel errors and propagated errors have a negative impact on the accuracy of the performance estimation. This bevahior was also addressed shortly in Section IV-B of \cite{SCFlip17-jour}. On the other hand, we show that there is a way to estimate the performance for $\omega>0$ accurately for Rate-1 and SPC nodes.

Based on (\ref{eqn:FERtheory}), the FER for a Rate-1 node under SC decoding can be derived using its top-node LLRs rather than its leaf node LLRs, by exploiting the fact that there are no frozen (parity) bits involved \cite{ProgressiveBitFlip_IEEEAccess2018}. This means that, the FER computation of a Rate-1 node can also be computed using the top-node LLRs instead of using its leaf node LLRs. As such, for a Rate-1 node at stage $S$ and of size $N_v$, (\ref{eqn:FERtheory}) can be adapted into
\begin{equation}\label{eqn:FERtheoryRate1_SC}
\text{FER}_{\text{Rate-1,SC}} = 1- (1-\pi^{S})^{N_v}\text{,}
\end{equation}
where $\pi^{S}$ is derived by substituting the mean value at stage $S$ ($\mu^S$) into (\ref{eqn:pi}). Note that, this derivation is possible because all variables at the top-level of a node have the same mean and variance. Similarly, we can derive the FER for higher error orders for Rate-1 nodes. Unlike the case in SC, the theoretical FER for Rate-1 nodes is accurate if it is calculated using their top-node LLRs, because channel-induced errors at top-level indices does not yield error propagation. In other words, the independent top-node errors in Rate-1 nodes is the key for obtaining accurate theroretical FER calculation. For instance, for a Rate-1 node of size $N_v$ and at stage $S$, the probability of a channel-induced error at a top-node index is $\pi^{S}$. To compute $\text{Pr}(e_1)$, we must ensure that all other indices are error-free ($(1-\pi^{S})^{N_v-1}$), and know that the channel error can occur in $N_v$ different indices:
\begin{equation*}%\label{eqn:FERtheoryRate1_e1}
\text{Pr}(e_1) = N_v \times \pi^{S}  (1-\pi^{S})^{N_v-1}\text{.}
\end{equation*}
  
The FER for a Rate-1 node with correcting one error at the top-level can be expressed as $1-\text{Pr}(e_{0})-\text{Pr}(e_1)$:
\begin{equation*}%\label{eqn:FERtheoryRate1_w1}
\text{FER}_{\text{Rate-1,}\omega=1} = 1- \big( (1-\pi^{S})^{N_v}\big) - \big(N_v \times \pi^{S}  (1-\pi^{S})^{N_v-1}\big)\text{.}
\end{equation*}

To generalize, the probability of error for a Rate-1 node of size $N_v$ at a target error order $\omega$ can be expressed as
\begin{equation}\label{eqn:FERtheoryRate1_w_short}
\text{FER}_{\text{Rate-1,}\omega} = 1- \sum_{i=0}^{\omega} \text{Pr}(e_i)\text{.}
\end{equation}
%Following the mathematical property $\sum_{k=0}^{N} {\binom{N}{k}} x^{k} (1-x)^{N-k} = 1$, (\ref{eqn:FERtheoryRate1_w_short}) can be expanded into
which can be expanded into
\begin{equation}\label{eqn:FERtheoryRate1_w}
\text{FER}_{\text{Rate-1,}\omega} = 1- \sum_{i=0}^{\omega} {\binom{N_v}{i}} \times \big{(}\pi^{S}\big{)}^i  (1-\pi^{S})^{N_v-i}\text{.}
\end{equation}

As the theoretical achievable FER limit for Rate-1 nodes is identified as a function of $N_v$, $\omega$ and $\mu^S$, we now attempt to approximate this with an artificially reduced size. We claim that the Rate-1 top-node indices with the lowest LLR magnitudes become far more susceptible to incur errors than others. Since the lowest LLR magnitudes at a parent node are propagated to its left child nodes using (\ref{eqn:alphaleft}), the FER of a Rate-1 node can be approximated by using its left child nodes instead of the root node. For example, the lowest LLR magnitude at the top of the Rate-1 node is also found at the leftmost leaf node ($S=0$), with a different mean value $\mu^0$. Similarly, the two lowest LLR magnitudes at the top of the Rate-1 node are also found at the leftmost node at $S=1$ with mean value $\mu^1$. Accordingly, we use the size and mean values of the left nodes of a Rate-1 node to obtain an approximated achievable FER limit to the original one in (\ref{eqn:FERtheoryRate1_w}). 

Let $\delta$ denote the number of indices within a \textit{search span} of a node which comprises the lowest LLR magnitudes at a Rate-1 node, where $N_v \geq \delta \geq \omega$. If $\delta < \omega$, then the number of errors cannot fit within the search span and a failed decoding is guaranteed. Accordingly, the achievable FER can be approximated as
\begin{equation}\label{eqn:FERtheoryRate1_w_approx}
\text{FER}_{\text{Rate-1,}\omega} \approx 1- \sum_{i=0}^{\omega} {\binom{\delta}{i}} \big{(}\pi^{*}\big{)}^i  (1-\pi^{*})^{\delta-i}
\end{equation}
where $\pi^{\ast}$ is the $\pi$ value calculated by the mean of the leftmost child at stage $S=\log_2\delta$. It should be noted that a similar study has been carried out in \cite{ProgressiveBitFlip_IEEEAccess2018} but it is limited to $\omega=0$ and $\delta=1$, which cannot be used for higher order FER approximations. 

\begin{figure}[t]
  \centering
  \input{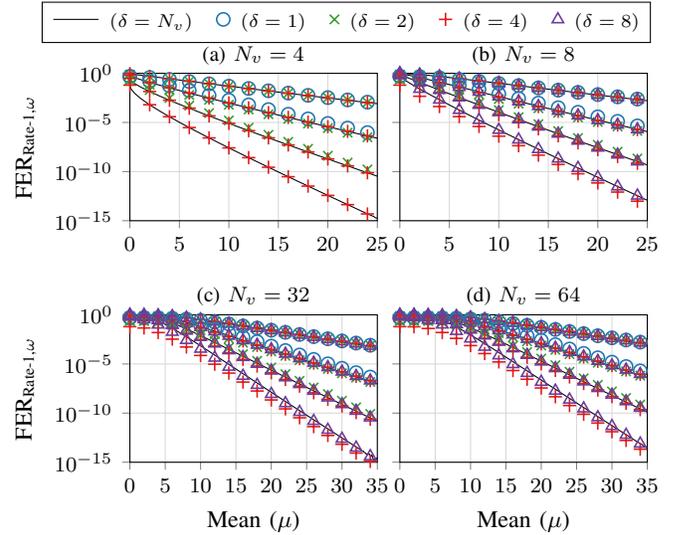}
  \vspace{-2em}
  \caption{\label{fig:Rate1curves} Theoretical achievable FER limits (\ref{eqn:FERtheoryRate1_w}) for Rate-1 nodes for $\omega \in \{0,1,2,3\}$ and different node sizes, compared to their approximations (\ref{eqn:FERtheoryRate1_w_approx}) using $\delta \in \{1,2,4,8\}$.}
\end{figure}

Fig.~\ref{fig:Rate1curves} depicts the exact achievable FER for Rate-1 nodes (\ref{eqn:FERtheoryRate1_w}) compared to the approximated versions (\ref{eqn:FERtheoryRate1_w_approx}) using four different $\delta$ values, four different node sizes and four error orders. It can be observed that, the approximations follow a closer trend to the original $\text{FER}_{\text{Rate-1,}\omega}$ with increasing $\delta$. At large node sizes and higher error orders, the approximations begin to diverge from the original achievable FER but they may still be considered within an acceptable domain, at a wide range of mean values. A reduced bit-flipping search span will be introduced for Rate-1 nodes based on the presented approximations at Section~\ref{sec:arch}.

\subsection{Reducing the Search Span in SPC Nodes}\label{sec:reducedsearch:spc}

At the top-level of an SPC node, all indices but one carry information and one bit carry the parity of all the other indices. This allows the SPC node to correct  one erroneous index \textit{naturally}, but certain conditions must be met. These conditions can be described as two probabilistic events: the probability of only one bit estimation is incorrect, and the probability of the incorrect bit containing the lowest absolute LLR value among all indices. If the incorrect bit does not have the lowest absolute LLR, then the correction mechanism of the SPC node causes an additional erroneous index instead of the correction. We denote the probability of a \textit{naturally correctable error} with $\text{Pr}(e^{*}_{1})$. %Since $\text{Pr}(e^{*}_{1})$ initially contains a single error, it is a subset of $\text{Pr}(e_{1})$. 

\begin{figure}
  \centering
  \scalebox{0.80}{
  %!TEX root = ../FastDSCF.tex
\begin{tikzpicture}
\begin{axis}[
  no markers, domain=-5:10, samples=100,
axis y line*=middle,
      axis x line*=bottom,
  xlabel=\empty, ylabel=\empty,
  every axis y label/.style={at=(current axis.above origin),anchor=south},
  every axis x label/.style={at=(current axis.right of origin),anchor=west},
  height=5cm, width=12cm,
  xtick={\empty}, ytick=\empty,
  enlargelimits=false, clip=false, axis on top,
  grid = major
  ]
  \addplot [fill=cyan!20, draw=none, domain=-5:0] {gauss(2.5,2)} \closedcycle;
  \addplot [fill=magenta!20, draw=none, domain=0.75:10] {gauss(2.5,2)} \closedcycle;
  \addplot [very thick,cyan!50!black] {gauss(2.5,2)};
  \node [below] at (axis cs:-5,0) {\footnotesize $-\infty$};
  \node [below] at (axis cs:10,0) {\footnotesize $\infty$};
  \node [below] at (axis cs:0,0) {\footnotesize $0$};

  \draw [dashed, color=black!50!white, ->](axis cs:-0.75,0.053) --  (axis cs:-1.25,0.080);
  %\node [above] at (axis cs:-1.25,0.080) {\small $\text{Pr}(L = x)$};
  \node [above] at (axis cs:-1.25,0.080) {\small $f(x;\mu,\sigma)$};
%%%%%%%%%%%%%%
  \draw [](axis cs:-0.75,0) --  (axis cs:-0.75,0.053);
  \draw [fill](axis cs:-0.75,0.053) circle (0.05cm);
  \node [below] at (axis cs:-0.75,0) {\footnotesize $x$};

  \draw [dashed, color=cyan!50!black, ->](axis cs:-1.25,0.020) --  (axis cs:-2.25,0.050);
  \node [above, color=cyan!50!black] at (axis cs:-2.75,0.045) {\small $Q(\frac{\mu}{\sigma})$};
%%%%%%%%%%%%%%
  \draw [](axis cs:0.75,0) --  (axis cs:0.75,0.136);
  \draw [fill](axis cs:0.75,0.136) circle (0.05cm);
  \node [below] at (axis cs:0.75,0) {\footnotesize $-x$};

  \draw [dashed, color=magenta!50!black, ->](axis cs:4.00,0.100) --  (axis cs:5.00,0.150);
  \node [above, color=magenta!50!black] at (axis cs:5.40,0.150) {\small $\text{Pr}(L \geq -x)$};
%%%%%%%%%%%%%%
  \draw [dashed, color=black!50!white](axis cs:2.5,0) --  (axis cs:2.5,0.200);
  \node [below] at (axis cs:2.5,0) {\footnotesize $\mu$};

%\draw [yshift=-0.6cm, latex-latex](axis cs:4,0) -- node [fill=white] {$1.96\sigma$} (axis cs:5.96,0);
\end{axis}

\end{tikzpicture}}
  \vspace{-2em}
  \caption{A Gaussian probability density function highlighting the Q-function, the probability of a negative value $x$, and the probability of a value equal to or larger than $-x$.}
  \label{fig:gaussian}
\end{figure}
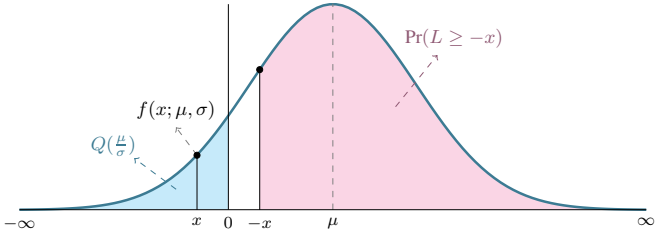

The associated events are visualized in the Gaussian probability density function in Fig.~\ref{fig:gaussian}, and explained next. Consider an SPC node of size $N_v$ at stage $S$ with top-level LLR values $\boldsymbol{L}^S$. Consider an all-zero codeword, a BPSK modulation and the AWGN channel. The probability of an SPC index $i$ is erroneous and naturally correctable can be described as the probability of index $i$ containing the LLR value of $L^S_i = x$ ($x<0$), and $\abs{x}$ is the smallest LLR magnitude at the top-level:
\begin{equation}\label{eqn:SPC_e1_fixable_example}
\text{Pr}(e^{*}_{1}, L^S_i=x) = \text{Pr}(L^S_i=x ~\cap~ \{\boldsymbol{L}^S \setminus L^S_i\} \geq -x )\text{.}
\end{equation}
%where $x$ is a negative LLR value at the top-level SPC index $i$.
The first event in (\ref{eqn:SPC_e1_fixable_example}) can be approximated using the normal probability density function:
\begin{equation}
f(x;\mu,\sigma) = e^{\frac{-(\mu-x)^2}{2\sigma^2}}/\sqrt{2\mathit{\pi}}\sigma \text{~,~} x \in \mathbb{R} \text{.}
\end{equation}
On the other hand, the second event can be described in terms of a Q-function with mean value shifted by $x$. Finally, since the two described events are statistically independent, we can apply $\text{Pr}(A \cap B) = \text{Pr}(A) \times \text{Pr}(B)$; leading to
\begin{equation}
\text{Pr}(e^{*}_{1} ,  L^S_i=x) = f(x;\mu,\sigma) \times \Big{[} 1-Q\Big{(}\frac{\mu+x}{\sigma}\Big{)} \Big{]}^{N_v-1} \text{.}
\end{equation}
Finally, taking into account that the described event in (\ref{eqn:SPC_e1_fixable_example}) may occur at any index, and for any $x<0$, we get
\begin{equation}\label{eqn:SPC_e1_fixable}
\text{Pr}(e^{*}_{1}) = {\binom{N_v}{1}} \int_{-\infty}^{0^-} f(x;\mu,\sigma) \Big{[} 1-Q\Big{(}\frac{\mu+x}{\sigma}\Big{)} \Big{]}^{N_v-1}~dx\text{.}
\end{equation}

Accordingly, the theoretical FER for an SPC node under SC decoding can be computed as:
\begin{equation}\label{eqn:FERtheorySPC_SC}
\text{FER}_{\text{SPC},\omega=0} = 1-\text{Pr}(e_{0})-\text{Pr}(e^{*}_{1})\text{,}
\end{equation}
where $\text{Pr}(e_{0})$ and $\text{Pr}(e^{*}_{1})$ can be substituted from (\ref{eqn:FER_SC_complement}) and (\ref{eqn:SPC_e1_fixable}). 

Now we show how to extend this computation for SPC nodes with higher error orders. Recall that two SPC indices are corrected per attempted error order. Hence, for $\omega=1$, two channel-induced errors are corrected, and a third error if it is correctable by the parity check ($\text{Pr}(e^{*}_{3})$). For the all-zero codeword scenario, given that an SPC index $i$ with an LLR value of $L_i=x$ ($x<0$), the remaining two erroneous indices must have lower values than $x$ and all other indices must have higher values than $-x$ so that index $i$ becomes correctable by the parity check. Considering all combinations of the described scenario and integrating for the range of $x$ yields
\begin{align}\label{eqn:SPC_e3_fixable}
\text{Pr}(e^{*}_{3}) =&~ 
   \int_{-\infty}^{0^-} {\binom{N_v}{1}} f(x;\mu,\sigma) {\binom{N_v-1}{2}} Q\Big{(}\frac{\mu-x}{\sigma}\Big{)}^2 \notag\\ & \times \Big{[} 1-Q\Big{(}\frac{\mu+x}{\sigma}\Big{)} \Big{]}^{N_v-3}~dx\text{.}
\end{align}

Following (\ref{eqn:FERtheoryRate1_w_short}), (\ref{eqn:FERtheorySPC_SC}) and (\ref{eqn:SPC_e3_fixable}) the theoretical FER of SPC nodes for any error order can be generalized as follows:
\begin{equation}\label{eqn:FERtheorySPC_w_short}
\text{FER}_{\text{SPC,}\omega} = 1- \sum_{i=0}^{2\omega} \text{Pr}(e_i) - \text{Pr}(e^{*}_{2\omega+1})\text{,}
\end{equation}
where
\begin{align}\label{eqn:SPC_e2w+1_fixable}
\text{Pr}(e^{*}_{2\omega+1}) = &~
    \int_{-\infty}^{0^-} {\binom{N_v}{1}} f(x;\mu,\sigma) {\binom{N_v-1}{2\omega}} Q\Big{(}\frac{\mu-x}{\sigma}\Big{)}^{2\omega} \notag\\ & \times \Big{[} 1-Q\Big{(}\frac{\mu+x}{\sigma}\Big{)} \Big{]}^{N_v-2\omega-1}~dx\text{.}
\end{align}

Our next and final goal is to approximate to the theoretical FER trend lines of SPC nodes with a limited search span. To do this, we perform a similar approach to the study in Rate-1 nodes: A search span that comprises $\delta$ lowest LLR magnitudes at an SPC node, where $N_v \geq \delta \geq \text{max}(2,2\omega+1)$, $\delta \in \mathbb{Z}^{+}$ is defined for SPC nodes. If $\delta < \text{max}(2,2\omega+1)$, then the number of errors cannot fit within the search span and a failed decoding is guaranteed. Accordingly, the achievable FER for SPC nodes within a limited search span can be expressed as 
\begin{align}\label{eqn:FERtheorySPC_w_approx}
\text{FER}_{\text{SPC,}\omega} &~ \approx 1- \sum_{i=0}^{2\omega} {\binom{\delta}{i}} \big{(}\pi^{*}\big{)}^i  (1-\pi^{*})^{\delta-i} \notag\\ & -  \int_{-\infty}^{0^-} {\binom{\delta}{1}} f(x;\mu^{*},\sigma^{*}) {\binom{\delta-1}{2\omega}} Q\Big{(}\frac{\mu^{*}-x}{\sigma^{*}}\Big{)}^{2\omega} \notag\\ & \times \Big{[} 1-Q\Big{(}\frac{\mu^{*}+x}{\sigma^{*}}\Big{)} \Big{]}^{\delta-2\omega-1}~dx\text{.}
\end{align}
where $\mu^{*}$ is obtained from the leftmost node at stage $S=\log_2\delta$, and $\pi^{*}$, and $\sigma^{*}$ are calculated using $\mu^{*}$.

\begin{figure}[t]
  \centering
  \begin{tikzpicture}

    \begin{groupplot}[
        group style={group name=group, group size= 2 by 2, horizontal sep=.3cm, vertical sep=1.25cm},
        footnotesize, width=\columnwidth, height=0.40\columnwidth,
        ymode=log,
        xmin = -5, xmax=5,
        ymin = 1.0e-15, ymax=1,
        xlabel={Mean ($\mu$)},
        width=0.5*\columnwidth,
        grid=both, grid style={gray!30},
        tick align=outside, tickpos=left, 
        ]

%w1
\nextgroupplot[ylabel={$\text{FER}_{\text{SPC,}\omega}$}, xlabel={}, width = .55\columnwidth, xtick={0, 5, ..., 25}, xmin=0, xmax=25]

\addplot[color=black, thin] table [x=MEAN,y=FER] {figures/SPCcurves/N4/w0}; \label{gp:refs}
\addplot[color=black, thin] table [x=MEAN,y=FER] {figures/SPCcurves/N4/w1}; 

\addplot[color=Paired-3, semithick, only marks,mark repeat=20, mark size = 2.5pt, mark options={solid}, mark=x] table [x=MEAN,y=FER] {figures/SPCcurves/N4/w0_s2};\label{gp:s2s}
\addplot[color=Paired-5, semithick, only marks,mark repeat=20, mark size = 2.5pt, mark options={solid}, mark=+] table [x=MEAN,y=FER] {figures/SPCcurves/N4/w0};\label{gp:s4s}
%\addplot[color=Paired-3, semithick, only marks,mark repeat=20, mark size = 2.5pt, mark options={solid}, mark=x] table [x=MEAN,y=FER] {figures/SPCcurves/N4/w1_s2};
\addplot[color=Paired-5, semithick, only marks,mark repeat=20, mark size = 2.5pt, mark options={solid}, mark=+] table [x=MEAN,y=FER] {figures/SPCcurves/N4/w1};%s4

\coordinate (top) at (rel axis cs:0,0);

\nextgroupplot[yticklabels={,,}, xlabel={}, width = .55\columnwidth, xtick={0, 5, ..., 25}, xmin=0, xmax=25]

\addplot[color=black, thin] table [x=MEAN,y=FER] {figures/SPCcurves/N8/w0};
\addplot[color=black, thin] table [x=MEAN,y=FER] {figures/SPCcurves/N8/w1}; 
\addplot[color=black, thin] table [x=MEAN,y=FER] {figures/SPCcurves/N8/w2}; 

\addplot[color=Paired-3, semithick, only marks,mark repeat=20, mark size = 2.5pt, mark options={solid}, mark=x] table [x=MEAN,y=FER] {figures/SPCcurves/N8/w0_s2};
\addplot[color=Paired-5, semithick, only marks,mark repeat=20, mark size = 2.5pt, mark options={solid}, mark=+] table [x=MEAN,y=FER] {figures/SPCcurves/N8/w0_s4};
\addplot[color=Paired-9, semithick, only marks,mark repeat=20, mark size = 2.5pt, mark options={solid}, mark=triangle] table [x=MEAN,y=FER] {figures/SPCcurves/N8/w0};\label{gp:s8s}

% \addplot[color=Paired-3, semithick, only marks,mark repeat=20, mark size = 2.5pt, mark options={solid}, mark=x] table [x=MEAN,y=FER] {figures/SPCcurves/N8/w1_s2};
\addplot[color=Paired-5, semithick, only marks,mark repeat=20, mark size = 2.5pt, mark options={solid}, mark=+] table [x=MEAN,y=FER] {figures/SPCcurves/N8/w1_s4};
\addplot[color=Paired-9, semithick, only marks,mark repeat=20, mark size = 2.5pt, mark options={solid}, mark=triangle] table [x=MEAN,y=FER] {figures/SPCcurves/N8/w1};

% \addplot[color=Paired-3, semithick, only marks,mark repeat=20, mark size = 2.5pt, mark options={solid}, mark=x] table [x=MEAN,y=FER] {figures/SPCcurves/N8/w2_s2};
% \addplot[color=Paired-5, semithick, only marks,mark repeat=20, mark size = 2.5pt, mark options={solid}, mark=+] table [x=MEAN,y=FER] {figures/SPCcurves/N8/w2_s4};
\addplot[color=Paired-9, semithick, only marks,mark repeat=20, mark size = 2.5pt, mark options={solid}, mark=triangle] table [x=MEAN,y=FER] {figures/SPCcurves/N8/w2};

\coordinate (top) at (rel axis cs:1,0);

%w2
\nextgroupplot[ylabel={$\text{FER}_{\text{SPC,}\omega}$}, width = .55\columnwidth,  xtick={0, 5, ..., 35}, xmin=0, xmax=35, ymin=1.0e-15]

\addplot[color=black, thin] table [x=MEAN,y=FER] {figures/SPCcurves/N32/w0};
\addplot[color=black, thin] table [x=MEAN,y=FER] {figures/SPCcurves/N32/w1}; 
\addplot[color=black, thin] table [x=MEAN,y=FER] {figures/SPCcurves/N32/w2}; 

\addplot[color=Paired-3, semithick, only marks,mark repeat=20, mark size = 2.5pt, mark options={solid}, mark=x] table [x=MEAN,y=FER] {figures/SPCcurves/N32/w0_s2};
\addplot[color=Paired-5, semithick, only marks,mark repeat=20, mark size = 2.5pt, mark options={solid}, mark=+] table [x=MEAN,y=FER] {figures/SPCcurves/N32/w0_s4};
\addplot[color=Paired-9, semithick, only marks,mark repeat=20, mark size = 2.5pt, mark options={solid}, mark=triangle] table [x=MEAN,y=FER] {figures/SPCcurves/N32/w0_s8};

% \addplot[color=Paired-3, semithick, only marks,mark repeat=20, mark size = 2.5pt, mark options={solid}, mark=x] table [x=MEAN,y=FER] {figures/SPCcurves/N32/w1_s2};
\addplot[color=Paired-5, semithick, only marks,mark repeat=20, mark size = 2.5pt, mark options={solid}, mark=+] table [x=MEAN,y=FER] {figures/SPCcurves/N32/w1_s4};
\addplot[color=Paired-9, semithick, only marks,mark repeat=20, mark size = 2.5pt, mark options={solid}, mark=triangle] table [x=MEAN,y=FER] {figures/SPCcurves/N32/w1_s8};

% \addplot[color=Paired-3, semithick, only marks,mark repeat=20, mark size = 2.5pt, mark options={solid}, mark=x] table [x=MEAN,y=FER] {figures/SPCcurves/N32/w2_s2};
% \addplot[color=Paired-5, semithick, only marks,mark repeat=20, mark size = 2.5pt, mark options={solid}, mark=+] table [x=MEAN,y=FER] {figures/SPCcurves/N32/w2_s4};
\addplot[color=Paired-9, semithick, only marks,mark repeat=20, mark size = 2.5pt, mark options={solid}, mark=triangle] table [x=MEAN,y=FER] {figures/SPCcurves/N32/w2_s8};

\coordinate (bot) at (rel axis cs:0,1);

\nextgroupplot[yticklabels={,,}, width = .55\columnwidth,  xtick={0, 5, ..., 35}, xmin=0, xmax=35, ymin=1.0e-15]

\addplot[color=black, thin] table [x=MEAN,y=FER] {figures/SPCcurves/N64/w0};
\addplot[color=black, thin] table [x=MEAN,y=FER] {figures/SPCcurves/N64/w1}; 
\addplot[color=black, thin] table [x=MEAN,y=FER] {figures/SPCcurves/N64/w2}; 

\addplot[color=Paired-3, semithick, only marks,mark repeat=20, mark size = 2.5pt, mark options={solid}, mark=x] table [x=MEAN,y=FER] {figures/SPCcurves/N64/w0_s2};
\addplot[color=Paired-5, semithick, only marks,mark repeat=20, mark size = 2.5pt, mark options={solid}, mark=+] table [x=MEAN,y=FER] {figures/SPCcurves/N64/w0_s4};
\addplot[color=Paired-9, semithick, only marks,mark repeat=20, mark size = 2.5pt, mark options={solid}, mark=triangle] table [x=MEAN,y=FER] {figures/SPCcurves/N64/w0_s8};

% \addplot[color=Paired-3, semithick, only marks,mark repeat=20, mark size = 2.5pt, mark options={solid}, mark=x] table [x=MEAN,y=FER] {figures/SPCcurves/N64/w1_s2};
\addplot[color=Paired-5, semithick, only marks,mark repeat=20, mark size = 2.5pt, mark options={solid}, mark=+] table [x=MEAN,y=FER] {figures/SPCcurves/N64/w1_s4};
\addplot[color=Paired-9, semithick, only marks,mark repeat=20, mark size = 2.5pt, mark options={solid}, mark=triangle] table [x=MEAN,y=FER] {figures/SPCcurves/N64/w1_s8};

% \addplot[color=Paired-3, semithick, only marks,mark repeat=20, mark size = 2.5pt, mark options={solid}, mark=x] table [x=MEAN,y=FER] {figures/SPCcurves/N64/w2_s2};
% \addplot[color=Paired-5, semithick, only marks,mark repeat=20, mark size = 2.5pt, mark options={solid}, mark=+] table [x=MEAN,y=FER] {figures/SPCcurves/N64/w2_s4};
\addplot[color=Paired-9, semithick, only marks,mark repeat=20, mark size = 2.5pt, mark options={solid}, mark=triangle] table [x=MEAN,y=FER] {figures/SPCcurves/N64/w2_s8};

\coordinate (bot) at (rel axis cs:1,1);

\end{groupplot}
    \node[above = 0.00cm of group c1r1.north] {\footnotesize(a) $N_v=4$};
    \node[above = 0.00cm of group c2r1.north] {\footnotesize(b) $N_v=8$};
    \node[above = 0.00cm of group c1r2.north] {\footnotesize(c) $N_v=32$};
    \node[above = 0.00cm of group c2r2.north] {\footnotesize(d) $N_v=64$};

    \path (top|-current bounding box.north) -- 
      coordinate(legendpos)
      (bot|-current bounding box.north);
    \matrix[
        matrix of nodes,
        anchor=south,
        draw,
        inner sep=0.2em,
        draw
      ]at([yshift=-.5ex, xshift=-3.6cm]legendpos)
      {
        \ref{gp:refs}& \footnotesize $(\delta=N_v)$ &[5pt]
        \ref{gp:s2s}&  \footnotesize $(\delta=2)$  &[5pt]
        \ref{gp:s4s}&  \footnotesize $(\delta=4)$ &[5pt]
        \ref{gp:s8s}&  \footnotesize $(\delta=8)$ \\};
\end{tikzpicture}
  \vspace{-2em}
  \caption{\label{fig:SPCcurves} Theoretical achievable FER limits (\ref{eqn:FERtheorySPC_w_short}) for SPC nodes for $\omega \in \{0,1,2\}$ and different node sizes, compared to their approximations (\ref{eqn:FERtheorySPC_w_approx}) using $\delta \in \{2,4,8\}$.}
\end{figure}
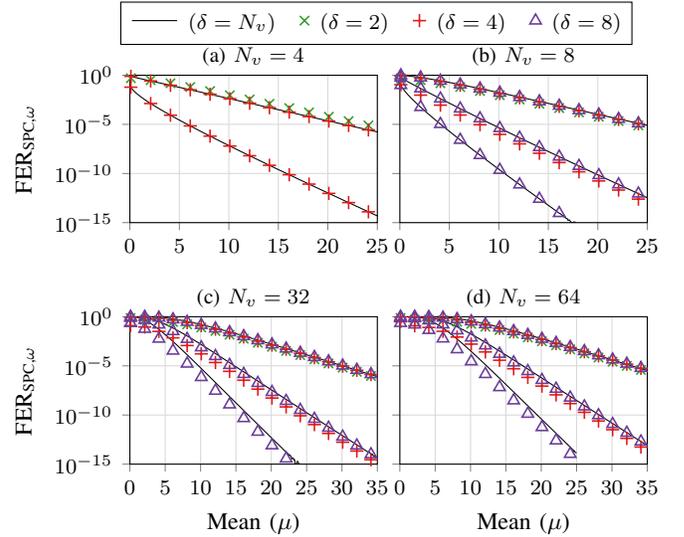

Fig.~\ref{fig:SPCcurves} depicts the exact achievable FER for SPC nodes (\ref{eqn:FERtheorySPC_w_short}) compared to the approximated version (\ref{eqn:FERtheorySPC_w_approx}) using three different $\delta$ values where applicable, four different node sizes and three error orders. Similar observations to the case of Rate-1 nodes can also be made for the SPC nodes: Considered approximations follow a close trend to the exact derivation, with higher $\delta$ values are in favor of a better approximation at the cost of higher search span. Similar to Rate-1 case, a reduced bit-flipping search span is created for SPC nodes based on the presented approximations at Section~\ref{sec:arch}.

\section{Hardware Architecture}\label{sec:arch}

The architecture for a typical SCF-based polar decoder comprises of the following main components: An SC decoder core where the decoding iterations are carried out, a CRC unit that works in parallel with the SC core to validate the output, and a sorter datapath to collect and regulate the bit-flipping information for possible additional decoding attempts. For this work, the SC decoder core is derived from the Fast-SCF decoder reported in \cite{FastSCF-TCAS-I}, which is based on a semi-parallel polar decoder architecture \cite{SPSC13} with a parallelization factor of $P_e=64$. Other than the branch operations, the modified decoder supports only Rep, Rate-1 and SPC instructions. Since Rate-0 nodes are always followed by a right branch (G) operation (\ref{eqn:alphaleft}), Rate-0 nodes are merged with right branch operations to reduce latency \cite{ErcanJSPS2018}. A highly-parallelized CRC processor for the polynomial $0\text{x}1021$ is implemented after \cite{CRC_generation} to validate the decoder output. 
 
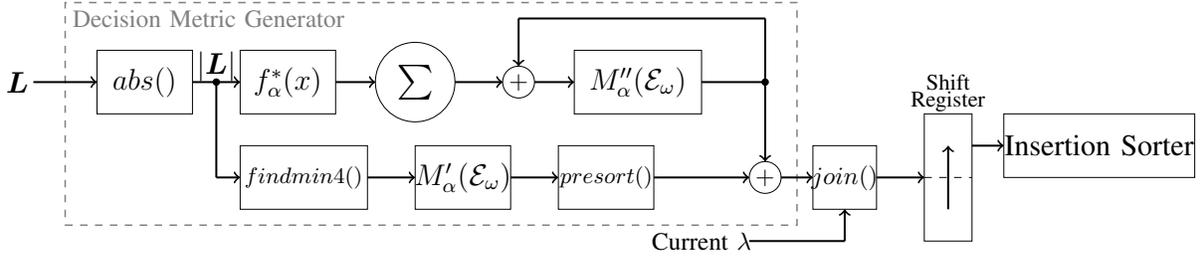
\begin{figure*}[t]
  \centering
  \scalebox{0.94}{%!TEX root = ../FastDSCF.tex

\begin{tikzpicture}[scale=.45]

\draw [dashed, color=black!50!white, semithick] (-1.00,3.50) rectangle (22.00,-3.50);
\node [color=black!50!white] at (3.50,3.00)  {Decision Metric Generator};

\draw [->,thick] (-2.00,1.00) -- (0.00,1.0);
\node [] at (-2.50,1.00)  {\large $\boldsymbol{L}$};

\draw (0.00,0.00) rectangle (3.00,2.00);
\node [] at (1.50,1.00)  {\large $abs()$};

\node [] at (3.75,1.50)  {\large $|\boldsymbol{L}|$};
\draw [->,thick] (3.00,1.00) -- (4.50,1.0);
\draw [fill=black] (3.75,1.00) circle [radius=0.1];
\draw [-,thick] (3.75,1.00) -- (3.75,-2.0);
\draw [->,thick] (3.75,-2.0) -- (4.50,-2.0);

\draw (4.50,0.00) rectangle (7.50,2.00);
\node [] at (6.00,1.00)  {\large $f^{*}_{\alpha}(x)$};

\draw [->,thick] (7.50,1.00) -- (8.75,1.0);

\draw [] (10.0,1.00) circle [radius=1.25];
\node [] at (10.0,1.00)  {\Large $\sum$};

\draw [->,thick] (11.25,1.00) -- (12.75,1.0);

\draw [] (13.25,1.00) circle [radius=0.50];
\node [] at (13.25,1.00)  { $+$};

\draw [->,thick] (13.75,1.00) -- (15.00,1.0);

\draw (15.0,0.00) rectangle (19.00,2.00);
\node [] at (17.0,1.00)  {\large $M''_{\alpha}(\mathcal{E}_{\omega})$};

\draw [-,thick] (19.00,1.00) -- (21.00,1.0);
\draw [fill=black] (21.00,1.00) circle [radius=0.1];
\draw [<-,thick] (21.00,-1.50) -- (21.00,3.0);
\draw [-,thick] (21.00,3.00) -- (13.25,3.0);
\draw [->,thick] (13.25,3.00) -- (13.25,1.5);

\draw (4.50,-3.00) rectangle (8.50,-1.00);
\node [] at (6.50,-2.00)  {\small $findmin4()$};
\draw [->,thick] (8.50,-2.00) -- (10.0,-2.0);
\draw (10.00,-3.00) rectangle (13.0,-1.00);
\node [] at (11.50,-2.00)  {\large $M'_{\alpha}(\mathcal{E}_{\omega})$};
\draw [->,thick] (13.00,-2.00) -- (14.5,-2.0);
\draw (14.50,-3.00) rectangle (17.5,-1.00);
\node [] at (16.0,-2.00)  {\small $presort()$};

\draw [->,thick] (17.50,-2.00) -- (20.5,-2.0);

\draw [] (21.00,-2.00) circle [radius=0.50];
\node [] at (21.00,-2.00)  {$+$};

\draw [->,thick] (21.50,-2.00) -- (22.5,-2.0);
\draw (22.50,-3.00) rectangle (24.5,-1.00);
\node [] at (23.50,-2.00)  {$join()$};

\node [] at (19.00,-4.00)  {Current $\lambda$};
\draw [-,thick] (20.50,-4.00) -- (23.50,-4.00);
\draw [->,thick] (23.50,-4.00) -- (23.5,-3.0);

\node [] at (26.75,1.00)  {\small Shift};
\node [] at (26.75,0.40)  {\small Register};
\draw [->,thick] (24.50,-2.00) -- (26.0,-2.0);
\draw (26.00,-4.00) rectangle (27.5,0.00);
\draw [dashed] (26.00,-2.00) -- (27.5,-2.0);
\draw [->,thick] (26.75,-3.00) -- (26.75,-1.0);

\draw [->,thick] (27.50,-1.00) -- (28.5,-1.0);
\draw (28.50,-2.00) rectangle (34.5,0.00);
\node [] at (31.50,-1.00)  {\large Insertion Sorter};

\end{tikzpicture}}
  \vspace{-1em}
  \caption{Sorter datapath for the proposed Fast-DSCF decoder architecture.}
  \label{fig:arch:FDSCFsorterdatapath}
  \vspace{-1em}
\end{figure*}

Compared to the state-of-the-art Fast-SCF decoder architectures, the main difference of the proposed Fast-DSCF architecture is on the sorter datapath, which includes generation of the decision metrics for bit-flipping candidates, followed by sorting and storing of the candidates. The architecture for the proposed sorter datapath is visualized in Fig.~\ref{fig:arch:FDSCFsorterdatapath}.

\vspace{-.9em}

\subsection{Metric Normalization and Quantization}\label{sec:arch:normalization}
The $M''_{\alpha}(\mathcal{E}_{\omega})$ component of the decision metric is accumulative as mentioned in Section~\ref{subsec:fastnodes}, which is not a desired property for quantization. When an accumulative component is quantized, it carries the risk of saturating over the course of the decoding, which disorganizes the sorted order of the bit-flipping indices and therefore may degrade the error-correction performance. One way to get around this problem is to increase the number of quantization bits at the cost of an increased latency, area and power consumption. Another way is to normalize the metric update by identifying and eliminating the computations that are performed more than once so that the risk of saturation could be minimized, as explained next.

Recall that the $M_{\alpha}(\mathcal{E}_{\omega})$ is computed as described in (\ref{eqn:dscfmetricMnew}), in which $M''_{\alpha}(\mathcal{E}_{\omega})$ is expressed as:
\begin{equation}
M''_{\alpha}(\mathcal{E}_{\omega}) = \sum_{j \in \mathcal{E}_{\omega-1}} | L^{0}[\mathcal{E}_{\omega-1}]_j | + \sum_{\substack{{j \leq i_{\omega}}\\ \forall j \in \mathcal{A}}} f_{\alpha}(| L^{0}[\mathcal{E}_{\omega-1}]_j |)\text{.}
\end{equation}
Note that, a part of this computation can be found in the preceding metric $M_{\alpha}(\mathcal{E}_{\omega-1})$. Following (\ref{eqn:dscfmetricM}) and (\ref{eqn:S}), $M_{\alpha}(\mathcal{E}_{\omega-1})$ can be expressed as:
\begin{equation}
M_{\alpha}(\mathcal{E}_{\omega-1}) = \sum_{j \in \mathcal{E}_{\omega-1}} | L^{0}[\mathcal{E}_{\omega-2}]_j | + \sum_{\substack{{j \leq i_{\omega-1}}\\ \forall j \in \mathcal{A}}} f_{\alpha}(| L^{0}[\mathcal{E}_{\omega-2}]_j |)\text{.}
\end{equation}
Given that $L^{0}[\mathcal{E}_{\omega-2}]_j = L^{0}[\mathcal{E}_{\omega-1}]_j$ for $j \leq i_{\omega-1}$, $M''_{\alpha}(\mathcal{E}_{\omega})$ can be normalized by $M_{\alpha}(\mathcal{E}_{\omega-1})$:
\begin{equation}
M''_{\alpha}(\mathcal{E}_{\omega}) - M_{\alpha}(\mathcal{E}_{\omega-1}) = \sum_{\substack{{i_{\omega-1} \leq j \leq i_{\omega}}\\ \forall j \in \mathcal{A}}} f_{\alpha}(| L^{0}[\mathcal{E}_{\omega-1}]_j |)\text{.}
\end{equation}
Therefore, the \textit{normalized} $M''_{\alpha}(\mathcal{E}_{\omega})$ can be initiated at $0$ at the beggining of any extra decoding attempt and remains unchanged until the last flipped index, and is updated only after the last flipped index as:
\begin{equation}\label{eqn:m2split_reduced}
M''_{\alpha}(\mathcal{E}_{\omega})_j \pluseq f_{\alpha}(| L^{0}[\mathcal{E}_{\omega-1}]_j|) \text{~if~} j \in \mathcal{A}, j > i_{\omega-1}.
\end{equation}
The normalization of $M''_{\alpha}(\mathcal{E}_{\omega})$ is not only in favor of quantization, but it also helps to reduce the computational effort by avoiding redundant calculations. Note that, this normalization procedure is extended to the computation of $M''_{\alpha}(\mathcal{E}_{\omega})$ at the special nodes that are detailed in (\ref{eqn:dscfmetricMnew2updateRep}), (\ref{eqn:dscfmetricMnew2updateRate1}), (\ref{eqn:dscfmetricMnew2updateSPC}) for the hardware implementation.

For quantization of Fast-DSCF decoding with $\omega=1$, $5$ bits and $6$ bits are set for the channel and internal LLRs, respectively, with $1$ bit reserved for the fractional part. A quantization of $5$ bits for the metric is shown to be sufficient to obtain a well-approximated performance to the floating-point decoding. 
On the other hand, these quantization schemes are not shown to be sufficient for higher order decoding: one extra bit is required for the fractional part of the LLRs which impacts channel and internal LLRs, and the metric. Moreover, another extra bit is required for the metric to sort the bit-flipping indices efficiently. Hence, $6$ and $7$ bits are set for the channel and internal LLRs, and $7$ bits are set for the metric quantization at $\omega>1$.

\vspace{-1em}

\subsection{Decision Metric Generation}\label{sec:arch:metric}
The $M'_{\alpha}(\mathcal{E}_{\omega})$ and $M''_{\alpha}(\mathcal{E}_{\omega})$ components of the decision metric are generated simultaneously, then summed to acquire the desired $M_{\alpha}$ for each bit-flipping candidate. During the execution of each special node, their top-node LLR values (shown as $\boldsymbol{L}$ in Fig.~\ref{fig:arch:FDSCFsorterdatapath}) are forwarded to the decision metric generator with the associated control signals, such as the instruction, parity information, last flipping location (if any), and the stage size.  The upper datapath in Fig.~\ref{fig:arch:FDSCFsorterdatapath} visualizes the signal flow on the generation of $M''_{\alpha}(\mathcal{E}_{\omega})$: the $f^{*}_{\alpha}(x)$ function from (\ref{eqn:fxconst}) is applied to the absolute top-node LLRs, which are then summed together and forwarded to the register that holds the current $M''_{\alpha}(\mathcal{E}_{\omega})$. The $M''_{\alpha}(\mathcal{E}_{\omega})$ component of the metric is updated based on the rule described in (\ref{eqn:m2split_reduced}), and the control signals are not shown for a simplified view.

Following the discussions on reducing the bit-flipping search span from Section~\ref{sec:reducedsearch}, the $M'_{\alpha}(\mathcal{E}_{\omega})$ component of the decision metric should only involve a predetermined number of indices. Following the studies in Fig.~\ref{fig:Rate1curves} and Fig.~\ref{fig:SPCcurves}, we reduce the search span for Rate-1 and SPC indices to $2$ and $4$, respectively. It can be seen from Fig.~\ref{fig:SPCcurves} that limiting the search span to $4$ for SPC nodes results in some performance degradation for larger node sizes at higher orders. On the other hand, increasing the search span for SPC nodes increases the complexity of the sorting process drastically. In order to minimize the performance degradation, the maximum  size for SPC nodes are limited for higher order decoding. Namely, the maximum SPC node sizes are set to $8$ and $4$ for $\omega=2$ and $\omega=3$, respectively.

The lower path in Fig.~\ref{fig:arch:FDSCFsorterdatapath} shows the $M'_{\alpha}(\mathcal{E}_{\omega})$ generation followed by the bit-flipping candidate generation. The $findmin4()$ function in Fig.~\ref{fig:arch:FDSCFsorterdatapath} finds the four indices with the minimum LLR magnitudes, with which $2$ and $\binom{4}{2} = 6$ bit-flipping candidates are generated with their $M'_{\alpha}(\mathcal{E}_{\omega})$ values for Rate-1 and SPC nodes, respectively. The generated candidates are pre-sorted ($presort()$ function in Fig.~\ref{fig:arch:FDSCFsorterdatapath}) which will greatly help with actual sorting process. The generated candidates are then assembled with the $M''_{\alpha}(\mathcal{E}_{\omega})$ component and with the current bit-flipping information ($join()$ function in Fig.~\ref{fig:arch:FDSCFsorterdatapath}). Hence, up to six bit-flipping candidates are generated that are forwarded to the sorter architecture. 

\vspace{-1em}

\subsection{Sorter Architecture}\label{sec:arch:sorter}

Since there is always at least one branch operation in between any two leaf node operations due to the sequential tree traversal \cite{ErcanJSPS2018}, the newly created sorting items can be processed in two clock cycles. Since up to six new sorting elements are created, up to three elements can be sorted at a clock cycle. This is achieved by inserting a dedicated shift register before the sorter architecture, that takes up to $6$ elements and sends $3$ elements to the sorter at a time. 

Let us denote a sorting element by $\lambda$, which contains the information of $\{\omega,M_{\alpha}(\mathcal{E}_{\omega}),\mathcal{E}_{\omega}\}$. The sorter architecture keeps an array of $\lambda$ items, $\boldsymbol{\lambda}$, with their metrics in increasing order. Let us further denote the newly generated sorting elements by $\boldsymbol{\lambda^{*}}=\{\lambda^{*}_{0},\lambda^{*}_{1},\lambda^{*}_{2}\}$. Note that the items in $\boldsymbol{\lambda^{*}}$ are also sorted in increasing order due to the $presort()$ function in Fig.~\ref{fig:arch:FDSCFsorterdatapath}. Since the length of $\boldsymbol{\lambda}$ is far greater than the size of $\boldsymbol{\lambda^{*}}$, together they form a \textit{nearly sorted} array, that needs to be fully sorted. The insertion sort method is an algorithm that is in favor of nearly sorted lists \cite{insertionSortRocks}, therefore we create an insertion sorter that is able to insert up to three new elements in a single clock cycle. Fig.~\ref{fig:arch:sorter_FDSCF} depicts the proposed insertion sorter architecture for Fast-DSCF decoding: The elements can shift back up to three places based on the places of the newly inserted values. The elements are also capable of shifting forward by one place, which is performed at the beginning of an additional decoding iteration. That way, the current bit-flipping information is always stored at $\lambda_{0}$. Based on the normalization procedure described in Section~\ref{sec:arch:normalization}, the metrics at each element of the sorter are normalized by the metric value of the current $\lambda$ (denoted as $\lambda_{0}\{M\}$) every time the sorter shifts its elements forward. 

\begin{figure}[t]
  \centering
  \scalebox{0.85}{%!TEX root = ../FastDSCF.tex

\begin{tikzpicture}[scale=.39]

\draw (0.00,0.00) rectangle (3.00,3.00);
\draw (4.50,0.00) rectangle (7.50,3.00);
\draw (9.00,0.00) rectangle (12.0,3.00);
\draw (13.50,0.00) rectangle (16.50,3.00);
\draw (18.00,0.00) rectangle (21.0,3.00);

\node [] at (1.50,1.50)  {\small $\boldsymbol{\lambda}_{0}$};
\node [] at (6.00,1.50)  {\small $\boldsymbol{\lambda}_{1}$};
\node [] at (10.5,1.50)  {\small $\boldsymbol{\lambda}_{2}$};
\node [] at (15.0,1.50)  {\small $\boldsymbol{\lambda}_{3}$};
\node [] at (19.5,1.50)  {\small $\boldsymbol{\lambda}_{l-1}$};

\node [] at (23.00, -2.00)  {\small $\boldsymbol{\lambda}^{*}$};

\draw [-,thick] (22.00,-2.50) -- (18.00,-2.50);
\draw [-,thick,dotted] (18.00,-2.50) -- (16.50,-2.50);
\draw [-,thick] (16.50,-2.50) -- (1.00,-2.50);
\draw [->,thick] (1.00,-2.50) -- (1.00,0.0);
\draw [->,thick] (5.50,-2.50) -- (5.50,0.0);
\draw [->,thick] (10.0,-2.50) -- (10.0,0.0);
\draw [->,thick] (14.5,-2.50) -- (14.5,0.0);
\draw [->,thick] (19.0,-2.50) -- (19.0,0.0);

\draw [-,thick] (22.00,-2.00) -- (18.00,-2.00);
\draw [-,thick,dotted] (18.00,-2.00) -- (16.50,-2.00);
\draw [-,thick] (16.50,-2.00) -- (6.00,-2.00);
\draw [->,thick] (6.00,-2.00) -- (6.00,0.0);
\draw [->,thick] (10.5,-2.00) -- (10.5,0.0);
\draw [->,thick] (15.0,-2.00) -- (15.0,0.0);
\draw [->,thick] (19.5,-2.00) -- (19.5,0.0);

\draw [-,thick] (22.00,-1.50) -- (18.00,-1.50);
\draw [-,thick,dotted] (18.00,-1.50) -- (16.50,-1.50);
\draw [-,thick] (16.50,-1.50) -- (11.00,-1.50);

\draw [->,thick] (11.0,-1.50) -- (11.0,0.0);
\draw [->,thick] (15.5,-1.50) -- (15.5,0.0);
\draw [->,thick] (20.0,-1.50) -- (20.0,0.0);

\draw [->,thick] (3.00,1.50) -- (4.50,1.50);
\draw [->,thick] (7.50,1.50) -- (9.00,1.50);
\draw [->,thick] (12.0,1.50) -- (13.5,1.50);
\draw [->,thick,dotted] (16.5,1.50) -- (18.0,1.50);

\draw[->,thick]   (2.50,3.00)  .. controls +(1,1) and +(-1,1) .. node[above] {} (9.50,3.00);
\draw[->,thick]   (7.00,3.00)  .. controls +(1,1) and +(-1,1) .. node[above] {} (14.0,3.00);
\draw[->,thick,dotted]   (11.5,3.00)  .. controls +(1,1) and +(-1,1) .. node[above] {} (18.5,3.00);

\draw[->,thick]   (2.50,0.00)  .. controls +(1,-1.5) and +(-1,-1.5) .. node[above] {} (14.0,0.00);
\draw[->,thick,dotted]   (7.00,0.00)  .. controls +(1,-1.5) and +(-1,-1.5) .. node[above] {} (18.5,0.00);

\draw [->,thick] (0.00,1.50) -- (-2.50,1.50);

\draw [-] (-1.00,1.50) -- (-1.00,2.625);
\draw [] (-1.00,3.00) circle [radius=0.375];
\node [] at (-1.00,3.00)  {\small $-$};
\draw [-] (-1.00,3.375) -- (-1.00,6.00);

\node [rotate=90] at (-1.50,4.75)  {\scriptsize $-\lambda_{0}\{M\}$};

\draw [-] (-1.00,6.00) -- (15.00,6.00);

\draw [] (1.50,4.75) circle [radius=0.375];
\node [] at (1.50,4.75)  {\small $+$};
\draw [] (6.00,4.75) circle [radius=0.375];
\node [] at (6.00,4.75)  {\small $+$};
\draw [] (10.5,4.75) circle [radius=0.375];
\node [] at (10.5,4.75)  {\small $+$};
\draw [] (15.0,4.75) circle [radius=0.375];
\node [] at (15.0,4.75)  {\small $+$};

\draw [-] (1.50,6.00) -- (1.50,5.125);
\draw [-] (6.00,6.00) -- (6.00,5.125);
\draw [-] (10.5,6.00) -- (10.5,5.125);
\draw [-] (15.0,6.00) -- (15.0,5.125);

\draw [->,thick] (1.50,4.375) -- (1.50,3.00);
\draw [->,thick] (6.00,4.375) -- (6.00,3.00);
\draw [->,thick] (10.5,4.375) -- (10.5,3.00);
\draw [->,thick] (15.0,4.375) -- (15.0,3.00);

\draw[->,thick]   (5.50,3.00)  .. controls +(-1,1) and +(1,0) .. node[above] {} (1.875,4.75);
\draw[->,thick]   (10.0,3.00)  .. controls +(-1,1) and +(1,0) .. node[above] {} (6.375,4.75);
\draw[->,thick]   (14.5,3.00)  .. controls +(-1,1) and +(1,0) .. node[above] {} (10.875,4.75);
\draw[->,thick,dotted]   (19.0,3.00)  .. controls +(-1,1) and +(1,0) .. node[above] {} (15.375,4.75);

\end{tikzpicture}}
  \vspace{-1em}
  \caption{\label{fig:arch:sorter_FDSCF} Insertion sorter architecture.}
  \vspace{-1.75em}
\end{figure}

With the increasing error order, the size of the sorter architecture increases in two dimensions: the cardinality of $\mathcal{E}_{\omega}$ within the sorting element ($\lambda$) increases and therefore more bits are required to describe a flipping event, and the length of the sorter increases linearly with $T_{\text{max}}$. Hence, the complexity of the sorter increases dramatically with $\omega$, requiring a substantial portion of the overall decoding architecture. To illustrate, for the Fast-DSCF decoder with $\omega=1$ and $T_{\text{max}}=10$, assuming $5$ bits for metric quantization, $2$ bits for order information, $9$ bits for the special node and $6$ bits for each stored index, $300$ bits are required for the sorter. On the other hand, assuming the same quantization numbers, $5,100$ and $28,800$ bits are required for the sorter for $\omega=3$ with $T_{\text{max}}=400$ and $\omega=2$ with $T_{\text{max}}=100$, respectively. In other words, the sorter complexity for $\omega=3$ would be about $5$ times of the LLR memory, and about $60$ times of the partial sum memory. Therefore, it is essential to reduce the sorter complexity for higher error orders as much as possible. 

For SCF architectures that feature $\omega=1$ only and not higher order error-correction, the sorter length must match the $T_{\text{max}}$ since the bit-flipping indices are calculated only once during the initial decoding attempt. On the other hand, for architectures that feature higher-order error-correction, the last elements at the sorter are most likely to be shifted out when the sorter gets updated with the higher-order bit-flipping information. Inspired from this event, we propose a sorter length $l \leq T_{\text{max}}$. When $l = T_{\text{max}}$, the error-correction performance of the Fast-DSCF algorithm is preserved. When $l < T_{\text{max}}$, the original error-correction performance is not guaranteed; however, an opportunity is created to reduce the sorter length at the expense of a preferably negligible loss in error-correction performance. Accordingly, empirical studies with $\omega=2$ and $\omega=3$ have shown that setting the sorter length to $50\%$ of the $T_{\text{max}}$ value has minimal impact on error-correction performance, while it greatly helps with reducing the computational complexity of the decoder. 

\section{Results}\label{sec:results}

The following results for the proposed Fast-DSCF decoder uses all the simplifications, optimizations and approximations discussed throughout this work. The constant approximation from (\ref{eqn:fxconst}), and all the special node decoding techniques from (\ref{eqn:dscfmetricMnew1Rep})-(\ref{eqn:dscfmetricMnew2updateSPC}) are used. Furthermore, following the discussion in Section~\ref{sec:reducedsearch}, the search span for Rate-1 and SPC nodes are reduced to $2$ and $4$, respectively. The maximum node size for SPC nodes is set to $\{64,8,4\}$ for $\omega \in \{1,2,3\}$, respectively. The metric normalization and quantization schemes following Section~\ref{sec:arch:normalization} are employed. For both DSCF and Fast-DSCF decoders, $T_{\text{max}}$ values are set to $10$, $100$ and $400$ for $\omega \in \{1,2,3\}$, respectively. The sorter lengths of the Fast-DSCF decoder are set to $10$, $50$ and $200$ for $\omega \in \{1,2,3\}$, respectively. A FER value of $10^{-5}$ is targeted for comparison.

\subsection{Error-Correction Performance}\label{sec:results:perf}

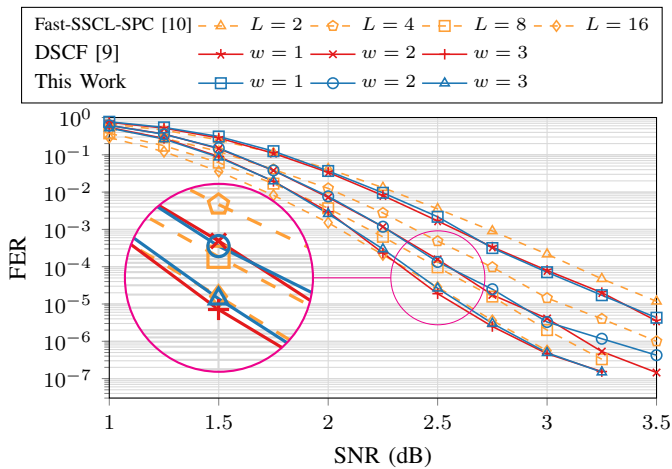
\begin{figure}[t]
  \centering
  %!TEX root = ../FastDSCF.tex

\begin{tikzpicture}[spy using outlines=
	{circle, magnification=2.0, connect spies}]

    \begin{semilogyaxis}[
            footnotesize, width=\columnwidth, height=.6\columnwidth,    
            xmin=1, xmax=3.50, xtick={1, 1.5,...,3.5},
            % ymin=1e-7,  
            ymax=1,
            xlabel=SNR (dB),%$\frac{E_s}{N_0} \text{~(dB)}$, %xlabel style={yshift=0.4em},
            ylabel=FER,  %ylabel style={yshift=-0.75em},
            grid=both, grid style={gray!30},
            tick align=outside, tickpos=left, 
            legend cell align={left},
            legend columns=5,
            legend style={at={(0.425,1.05)},anchor=south}],
        ]

\addlegendimage{empty legend}
%\addlegendentry{\scriptsize Fast-SSCL-SPC \cite{fastSSCL-TSP}}
\addlegendentry{\scriptsize Fast-SSCL-SPC \cite{fastSSCL-TSP}}

\addplot[
    color=Paired-7!50!Paired-8,
    dashed,
    mark options={solid},
    mark=triangle,
    semithick
]
table {
1.000  0.67576758
1.250  0.45994599
1.500  0.25042504
1.750  0.11211121
2.000  0.04204507
2.250  0.01320341
2.500  0.00355290
2.750  0.00090700
3.000  0.00021638
3.250  0.00004740
3.500  0.00001180
};
\addlegendentry{\scriptsize $L=2$}%

\addplot[
    color=Paired-7!50!Paired-8,
    dashed,
    mark options={solid},
    mark=pentagon,
    semithick
]
table {
1.000 0.50085009
1.250 0.27902790
1.500 0.12301230
1.750 0.04051864
2.000 0.01259129
2.250 0.00276765
2.500 0.00048413
2.750 0.00009700
3.000 0.00001440
3.250 0.00000400
3.500 0.00000100
};
\addlegendentry{\scriptsize $L=4$}%

\addplot[
    color=Paired-7!50!Paired-8,
    dashed,
    mark options={solid},
    mark=square,
    semithick
]
table {
1.000 0.37603760
1.250 0.17561756
1.500 0.06160616
1.750 0.01688961
2.000 0.00368979
2.250 0.00063292
2.500 0.00009640
2.750 0.00001580
3.000 0.00000200
3.250 0.00000033
};
\addlegendentry{\scriptsize $L=8$}%

\addplot[
    color=Paired-7!50!Paired-8,
    dashed,
    mark options={solid},
    mark=diamond,
    semithick
]
table {
1.000 0.29082908
1.250 0.12131213
1.500 0.03581662
1.750 0.00823655
2.000 0.00151682
2.250 0.00020811
2.500 0.00002840
2.750 0.00000360
3.000 0.00000056
};
\addlegendentry{\scriptsize $L=16$}%

%w=2,t=40

\addlegendimage{empty legend}
\addlegendentry{DSCF \cite{SCFlip17-jour}}

\addplot[
    color=Paired-5,
    mark=star,
    semithick,
]
table {
1.00    7.42900e-01
1.25    5.21900e-01
1.50    2.82800e-01
1.75    1.10100e-01
2.00    3.38000e-02
2.25    8.10000e-03
2.50    1.68816e-03
2.75    3.36945e-04
3.00    7.92511e-05
3.25    2.01765e-05
3.50    3.51435e-06
};
\addlegendentry{\scriptsize $w=1$}

\addplot[
    color=Paired-5,
    mark=x,
    semithick,
]
table {
1.00	6.14800e-01
1.25	3.61200e-01
1.50	1.46100e-01
1.75	3.80000e-02
2.00	7.20000e-03
2.25	1.17272e-03
2.50	1.57300e-04
2.75	1.75323e-05
3.00	4.09013e-06
3.25	5.31359e-07
3.50    1.45093e-07
};
\addlegendentry{\scriptsize $w=2$}

%w=3,t=200
\addplot[
    color=Paired-5,
    mark=+,
    semithick
]
table {
1.00    5.40900e-01
1.25    2.75800e-01
1.50    9.03000e-02
1.75    1.94000e-02
2.00    2.97637e-03
2.25    2.32769e-04
2.50    1.89323e-05
2.75    2.49382e-06
3.00    4.63922e-07
3.25    1.54245e-07
};
\addlegendentry{\scriptsize $w=3$}

\addlegendimage{empty legend}
\addlegendentry{}

\addlegendimage{empty legend}
\addlegendentry{This Work}

\addplot[
    color=Paired-1,
    mark=square,
    semithick
]
table {
1.00    7.69600e-01
1.25    5.44200e-01
1.50    3.10300e-01
1.75    1.25200e-01
2.00    3.69000e-02
2.25    9.60000e-03
2.50    2.17950e-03
2.75    3.15918e-04
3.00    7.11453e-05
3.25    1.71957e-05
3.50    4.28197e-06
};
\addlegendentry{\scriptsize $w=1$}

\addplot[
    color=Paired-1,
    mark=o,
    semithick
]
table {
1.00 6.14600e-01
1.25 3.55200e-01
1.50 1.51500e-01
1.75 3.83000e-02
2.00 7.70000e-03
2.25 1.18209e-03
2.50 1.34157e-04
2.75 2.48948e-05
3.00 3.29184e-06
3.25 1.17025e-06
3.50 4.23950e-07
};
\addlegendentry{\scriptsize $w=2$}

\addplot[
    color=Paired-1,
    mark=triangle,
    semithick
]
table {
1.00 5.21500e-01
1.25 2.64800e-01
1.50 8.59000e-02
1.75 1.97000e-02
2.00 2.62591e-03
2.25 2.97057e-04
2.50 2.64217e-05
2.75 3.00366e-06
3.00 4.92056e-07
3.25 1.48587e-07
};
\addlegendentry{\scriptsize $w=3$}

\coordinate (spypoint) at (axis cs:2.5,5.0e-5);
\coordinate (magnifyglass) at (axis cs:1.50,5e-5);

\end{semilogyaxis}
\spy [magenta, height=2.5cm, width=2.5cm] on (spypoint)
   in node[fill=white] at (magnifyglass);
\end{tikzpicture}
  \vspace{-2em}
  \caption{\label{fig:results_FER}FER comparison of the proposed Fast-DSCF decoding against baseline DSCF and Fast-SSCL-SPC \cite{fastSSCL-TSP} decoders, using $PC(1024, 512)$ and $C =16$. The $T_{\text{max}}$ are set to $10,100,400$ for $\omega \in \{1,2,3\}$ for both DSCF and Fast-DSCF decoders. The Fast-SSCL-SPC and the Fast-DSCF decoders are quantized, whereas DSCF decoder is simulated using floating-point.}
  \vspace{-1.5em}
\end{figure}

Fig.~\ref{fig:results_FER} presents the error-correction performance of the proposed Fast-DSCF decoding, against baseline DSCF decoding from \cite{SCFlip17-jour} and CRC-aided Fast-SSCL-SPC (Fast-SSCL) decoding with $L \in \{2,4,8,16\}$ \cite{fastSSCL-TSP}, using $PC(1024,512)$ and $C=16$ from \cite{38.212}. Note that the Fast-DSCF is simulated using the quantization schemes presented in Section~\ref{sec:arch:normalization}, whereas the DSCF decoder is simulated using floating point numbers, and the Fast-SSCL decoder is quantized with the scheme presented in \cite{fastSSCL-TSP}. 

According to Fig.~\ref{fig:results_FER}, the Fast-DSCF decoder equipped with all the simplifications exhibits similar performance to the baseline DSCF decoder at all error orders and SNRs. It can be observed that the Fast-DSCF decoder has similar performance to DSCF even though the Fast-DSCF is quantized. At $\omega=2$ and high SNR, Fast-DSCF shows a slight performance loss compared to the DSCF; this is mostly due to the reduced search span in SPC nodes. This means that a better SPC node approximation could be required at FER values targeting beyond $10^{-6}$. 

At the target FER of $10^{-5}$, the proposed Fast-DSCF at $\omega=1$ depicts $0.18$ dB gain over Fast-SSCL-SPC with $L=2$. At $\omega=2$, Fast-DSCF is $0.21$ dB better than Fast-SSCL-SPC with $L=4$ and is only $0.08$ dB away from the Fast-SSCL-SPC performance with $L=8$. Compared to Fast-SSCL-SPC with $L=16$, proposed Fast-DSCF performs slightly better than Fast-SSCL-SPC, by $0.03$ dB. In the following comparison schemes, based on these performance results Fast-DSCF is compared against its Fast-SSCL-SPC counterparts that yield the closest error-correction performance at target FER$=10^{-5}$. That is, Fast-DSCF with $\omega=1,2,3$ are compared against Fast-SSCL-SPC (and other state-of-the-art SCL-based decoders) with $L=2,8,16$, respectively.

\subsection{Average Computational Complexity}\label{sec:results:iter}

\begin{figure}[t]
  %!TEX root = ../FastDSCF.tex

\begin{tikzpicture}
    \begin{semilogyaxis}[
            footnotesize, width=\columnwidth, height=0.60\columnwidth,    
            xmin=2.00, xmax=3.75,xtick={2.00,2.25,...,3.50},
            ytick={300,500,...,1500},
            log ticks with fixed point,
            % ymin=1e-7,  ymax=2,
            xlabel=SNR (dB),%$\frac{E_s}{N_0} \text{~(dB)}$, 
            ylabel={Avg. $\#$ of Clock Cycles},
            grid=both, grid style={gray!30},
            tick align=outside, tickpos=left, 
            legend columns =4,
            legend style={/pgfplots/on layer=axis tick labels},
            legend pos=north east],
        ]

\addlegendimage{empty legend}
\addlegendentry{\scriptsize Fast-SSCL-SPC \cite{fastSSCL-TSP} }

\addplot[
    color=Paired-1,
    mark=triangle,
    dashed,
    mark options={solid},
    semithick,
]
table {
1.50    321
1.75    321
2.00    321
2.25    321
2.50    321
2.75    321
3.00    321
3.25    321
3.50    321
3.75    321
};
\addlegendentry{\tiny $L=2$}

\addplot[
    color=Paired-3,
    mark=square,
    dashed,
    mark options={solid},
    semithick,
]
table {
1.50    487
1.75    487
2.00    487
2.25    487
2.50    487
2.75    487
3.00    487
3.25    487
3.50    487
3.75    487
};
\addlegendentry{\tiny $L=8$}

\addplot[
    color=Paired-5,
    mark=diamond,
    dashed,
    mark options={solid},
    semithick,
]
table {
1.50    571
1.75    571
2.00    571
2.25    571
2.50    571
2.75    571
3.00    571
3.25    571
3.50    571
3.75    571
};
\addlegendentry{\tiny $L=16$}

\addlegendimage{empty legend}
\addlegendentry{\scriptsize Fast-SSCL-SPC \cite{Kim_Park_FSSCLSPC} }

\addplot[
    color=Paired-1,
    mark=x,
    dashed,
    mark options={solid},
    semithick,
]
table {
1.50    3.66E+02
1.75    3.62E+02
2.00    3.59E+02
2.25    3.56E+02
2.50    3.53E+02
2.75    3.50E+02
3.00    3.48E+02
3.25    3.45E+02
3.50    3.42E+02
3.75    3.39E+02
};
\addlegendentry{\tiny $L=2$}

\addplot[
    color=Paired-3,
    mark=+,
    dashed,
    mark options={solid},
    semithick,
]
table {
1.50    5.31E+02
1.75    5.25E+02
2.00    5.21E+02
2.25    5.16E+02
2.50    5.11E+02
2.75    5.05E+02
3.00    5.00E+02
3.25    4.95E+02
3.50    4.91E+02
3.75    4.87E+02
};
\addlegendentry{\tiny $L=8$}

\addplot[
    color=Paired-5,
    mark=star,
    dashed,
    mark options={solid},
    semithick,
]
table {
1.50    6.14E+02
1.75    6.08E+02
2.00    6.01E+02
2.25    5.95E+02
2.50    5.89E+02
2.75    5.81E+02
3.00    5.75E+02
3.25    5.69E+02
3.50    5.64E+02
3.75    5.59E+02
};
\addlegendentry{\tiny $L=16$}

\addlegendimage{empty legend}
\addlegendentry{\scriptsize This Work}

\addplot[
    color=Paired-1,
    mark=o,
    semithick,
]
table {
1.50    1.62E+03
1.75    9.51E+02
2.00    5.86E+02
2.25    4.40E+02
2.50    3.92E+02
2.75    3.77E+02
3.00    3.73E+02
3.25    3.73E+02
3.50    3.73E+02
3.75    3.73E+02
};
\addlegendentry{\tiny $w=1$}

\addplot[
    color=Paired-3,
    mark=x,
    semithick,
]
table {
1.50    8.18E+03
1.75    2.80E+03
2.00    1.01E+03
2.25    5.22E+02
2.50    4.36E+02
2.75    4.19E+02
3.00    4.15E+02
3.25    4.11E+02
3.50    4.11E+02
3.75    4.11E+02
};
\addlegendentry{\tiny $w=2$}

\addplot[
    color=Paired-5,
    mark=pentagon,
    %ticklabel style= {/pgfplots/on layer=foreground},
    semithick,
]
table {
1.50    2.17E+04
1.75    5.90E+03
2.00    1.43E+03
2.25    6.03E+02
2.50    4.74E+02
2.75    4.56E+02
3.00    4.51E+02
3.25    4.51E+02
3.50    4.47E+02
3.75    4.47E+02
};
\addlegendentry{\tiny $w=3$}

\end{semilogyaxis}
\end{tikzpicture}
  \vspace{-2em}
  \caption{\label{fig:results_ITER}Comparison of average number of decoding steps for proposed simplified DSCF decoding and baseline DSCF decoding. $PC(1024, 512), C =16$.}
  \vspace{-1em}
\end{figure}
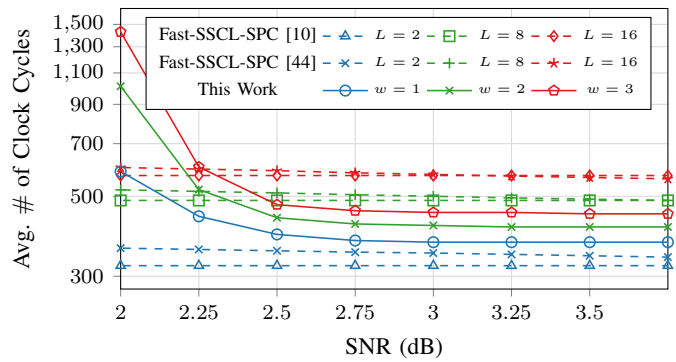

Fig.~\ref{fig:results_ITER} compares the average computational complexity of Fast-DSCF decoder against state-of-the-art Fast-SSCL-SPC decoders from \cite{fastSSCL-TSP} and \cite{Kim_Park_FSSCLSPC}, at medium-to-high SNR regime. The computational complexity is measured via the average number of cycles, which is obtained by taking the product of the average number of decoding iterations with the total number of clock cycles per iteration, for each decoder. It is essential to note that, the number of cycles are obtained with using the same polar code ($PC(1024,512)$) and using the same number of parallel processing elements ($P_e=64$) for all considered decoders. While latency of the Fast-SSCL-SPC decoder from \cite{fastSSCL-TSP} is fixed at each list size, the decoder from \cite{Kim_Park_FSSCLSPC} uses early termination logic and its latency is dependent on the SNR. The Fast-DSCF decoder has large computational complexity at low SNR regions, but saturates quickly around our FER performance of interest. 

At medium-to-high SNR regimes, the Fast-DSCF with $\omega=1$ requires up to $16\%$ more clock cycles than the its counterparts with $L=2$. On the other hand, for $\omega=2$ and $\omega=3$, Fast-DSCF requires $15.6\%$ and $21.7\%$ less cycles on average than the Fast-SSCL-SPC decoders with $L=8$ and $L=16$, respectively. Therefore, we can claim that the proposed Fast-DSCF decoding for higher order error correction requires less amount of operations on average compared to Fast-SSCL-SPC decoding, which makes it favorable for applications that require improved FER performance with less average computational complexity.

\begin{table*}[ht]
\centering
\caption{TSMC $65$nm CMOS implementation results for proposed Fast-DSCF architectures featuring $\omega \in \{1,2,3\}$ against equivalent state-of-the-art SCF and SCL decoders with $PC(1024,512)$.}
\label{tab:results}
\resizebox*{2.00\columnwidth}{!}{
\begin{tabular}{@{}cccccccccccc@{}}
\toprule
Base Algorithm                                   & \multicolumn{6}{c}{SCF}       & \multicolumn{5}{c}{SCL} \\
\cmidrule(l){2-7} \cmidrule(l){8-12}
                                                 & \multicolumn{3}{c}{This Work} & \cite{FastSCF-TCAS-I} & \cite{FTSCF_ICC20} & \cite{PolarBear}\textsuperscript{(a)}                  & \multicolumn{2}{c}{\cite{fastSSCL-TSP}} & \multicolumn{2}{c}{\cite{Xia_SCL_TSP}\textsuperscript{(a)}}                     & \cite{Fan_SCL_JSAC}\textsuperscript{(a)}             \\
\cmidrule(l){2-4} \cmidrule(l){5-5} \cmidrule(l){6-6} \cmidrule(l){7-7} \cmidrule(l){8-9} \cmidrule(l){10-11} \cmidrule(l){12-12}
 Methodology                                      & \multicolumn{3}{c}{Synthesis} & Synthesis             & Synthesis   & Silicon                    & \multicolumn{2}{c}{Synthesis}    & \multicolumn{2}{c}{Synthesis}                       & Synthesis                \\
 Technology (nm)                                  & \multicolumn{3}{c}{65}        & 65                    & 65          & 28                         & \multicolumn{2}{c}{65}           & \multicolumn{2}{c}{90}                              & 90                       \\
 Supply (V)                                       & \multicolumn{3}{c}{1.0}       & 1.0                   & 1.0         & 0.9                        & \multicolumn{2}{c}{-}            & \multicolumn{2}{c}{-}                               & -                        \\
 $P_e$                                            & \multicolumn{3}{c}{64}        & 64                    & 64          & -                          & \multicolumn{2}{c}{64}           & \multicolumn{2}{c}{-}                               & 64                       \\

 $\omega$ (SCF) / $L$ (SCL)                          & 1        & 2        & 3       & 1                     & 1           & 1                          & 2 & 8            & 8 & 16                                & 16                        \\
 \cmidrule(l){2-2} \cmidrule(l){3-3} \cmidrule(l){4-4} 
 \cmidrule(l){5-5} \cmidrule(l){6-6} \cmidrule(l){7-7} 
 \cmidrule(l){8-8} \cmidrule(l){9-9} \cmidrule(l){10-10} 
 \cmidrule(l){11-11} \cmidrule(l){12-12}  
 %List Size (L)                                    & N/A      & N/A      & N/A     & N/A                   & N/A         & N/A                        & 2               & 8              & 8                        & 16                       & 16                       \\
 
 $T_{\text{max}}$                                 & 10       & 100      & 400     & 20                    & 10          & 20                         & \multicolumn{2}{c}{N/A}            & \multicolumn{2}{c}{N/A}                               & N/A                        \\
 $Q_{\text{chn}}$,$Q_{\text{int}}$,$Q_{\text{m}}$ & 5,6,5    & 6,7,7    & 6,7,7   & 5,6,5                 & 5,6,5       & 6,6,8                      & \multicolumn{2}{c}{6,6,8}        & \multicolumn{2}{c}{6,6,8}                           & 6,6,8                    \\
  Frequency (MHz)                                  & 455      & 425      & 410     & 455                   & 480         & 145    & 885             & 722            & 770  & 676  & 911  \\
  Latency ($\mu$s)                                 & 0.82\textsuperscript{(b)}     & 0.97\textsuperscript{(b)}     & 1.11\textsuperscript{(b)}    & 0.68                  & 0.64        & 12.68  & 0.55            & 0.85           & 0.67 & 0.76 & 1.60 \\
  W.C. Latency ($\mu$s)                            & 9.02     & 49.7     & 447     & 14.23                 & 7.04        & 266.18 & 0.55            & 0.85           & 0.67 & 0.76 & 1.60 \\
  Average Coded T/P (Mbps)                         & 1248\textsuperscript{(b)}     & 1060\textsuperscript{(b)}     & 935\textsuperscript{(b)}     & 1511                  & 1595        & 80.8   & 1861            & 1198           & 1527 & 1340 & 637  \\
  Area (mm$^2$)                                    & 0.55     & 0.69     & 1.00    & 0.56                  & 0.49        & -                          & 1.05            & 3.98           & 2.37 & 4.21 & 3.90 \\
  Power Consumption (mW)                           & 118.01   & 136.35   & 201.61  & 83.44                 & -           & 51.30                      & - & -            & - & -                               & -                        \\
  Power Density (mW/mm$^2$)                        & 214.56   & 197.61   & 201.61  & 149.53                & -           & -                          & - & -            & - & -                               & -                        \\
  Area Efficiency (Gbps/mm$^2$)                    & 2.27     & 1.54     & 0.94    & 2.71                  & 3.25        & -                          & 1.78            & 0.30            & 0.64 & 0.32 & 0.16 \\
  Energy (pJ/info.bit)                             & 189.14   & 259.79   & 439.86  & 110.43                & -           & 1270   & - & -            & - & -                               & -                       \\
\bottomrule
\multicolumn{12}{l}{\textsuperscript{(a)} {\scriptsize Normalized for $65$nm CMOS technology and 1V supply, based on the scaling techniques from \cite{PolarBear}}.} \\
\multicolumn{12}{l}{\textsuperscript{(b)} {\scriptsize Average value at target FER$=10^{-5}$.}} \\
\end{tabular}}
%\vspace{-2.25em}
\end{table*}

\subsection{ASIC Synthesis Results}\label{sec:results:asic}

The proposed Fast-DSCF decoder has been implemented in VHDL, validated with test benches and synthesized using TSMC $65$nm CMOS technology node through Cadence Genus RTL compiler. To assure accuracy in our power measurements, switching activities from real test frames are extracted for the three architectures. The non-frozen bits for the test frames are generated using Bernoulli distribution equal probability.

Table~\ref{tab:results} presents the ASIC synthesis results of the Fast-DSCF decoder implemented separately for each considered error order, and compared against other available SCF-based decoders and best available SCL-based decoders. The results from \cite{PolarBear}, \cite{Xia_SCL_TSP} and \cite{Fan_SCL_JSAC} are scaled to $65$nm CMOS technology, based on the presented scaling techniques in \cite{PolarBear}. The latency and the throughput for this work are calculated based on their average values at target FER$=10^{-5}$. On the other hand, the worst-case latency values are also presented for a fair comparison. The quantization values are presented for the channel ($Q_{\text{chn}}$) and internal LLRs ($Q_{\text{int}}$) and for the metric ($Q_{\text{m}}$).

The latency and the average throughput of the Fast-DSCF decoder is different for each implementation; that is because their different quantization schemes lead to different operating frequencies, and different SPC node sizes lead to different number of clock cycles per decoding iteration. On the other hand, the power consumption and the area increases with $\omega$: this is mostly due to the increased complexity of the sorter. Specifically, the insertion sorter component consumes $2.2\%$ of the overall power consumption at $\omega=1$, its share increases to $16.1\%$ and $46.5\%$ for $\omega=2$ and $\omega=3$, respectively. 

The only available SCF-based decoders in the literature that report ASIC results target $\omega=1$ \cite{PolarBear,FastSCF-TCAS-I,FTSCF_ICC20}. According to Table~\ref{tab:results}, the former Fast-SCF \cite{FastSCF-TCAS-I} and Fast-TSCF \cite{FTSCF_ICC20} decoders provide a higher throughput than the proposed Fast-DSCF decoder at $\omega=1$. The main reason to this is that the former architectures use merged special node techniques (\textit{e.g.} Rep-SPC) that requires complex metric updates and bit-flips in our framework, and thus they are not implemented. Moreover, the Fast-TSCF decoder does not use a sorter, that allows for slightly higher throughput. The worst-case latency of the Fast-DSCF decoder is also more than that of Fast-TSCF, and it is mainly due to the merged special node techniques embedded in the Fast-TSCF. However, the worst-case latency of Fast-SCF is worse than the proposed scheme because it requires a larger $T_{\text{max}}$ value to reach a similar performance to Fast-DSCF decoder. On the other hand, the fast decoding techniques used in the Fast-DSCF ($\omega=1$) decoder yields $15.4\times$ more throughput than the regular SCF implementation from \cite{PolarBear}. Finally, compared to the Fast-SSCL-SPC decoder with $L=2$ \cite{fastSSCL-TSP}, the proposed decoder has $32\%$ less throughput but uses $1.9\times$ less area, leading to $27.5\%$ better area efficiency.

The Fast-DSCF decoder with $\omega=2$ demonstrates $11.5\%$ to $30.5\%$ less throughput than its SCL-based counterparts with $L=8$ \cite{fastSSCL-TSP,Xia_SCL_TSP}. On the other hand, the required area for the SCL-based decoders is $3.4\times$ to $5.8\times$ more than that of Fast-DSCF. Therefore, the Fast-DSCF decoder with $\omega=2$ is $2.4\times$ to $5.1\times$ more area-efficient than the reported fast SCL decoders. The Fast-DSCF decoder with $\omega=3$ reports $46.8\%$ more throughput than the SCL-based decoder in \cite{Fan_SCL_JSAC}, but is $30\%$ less than that of \cite{Xia_SCL_TSP} with $L=16$. On the other hand, similar to the $\omega=2$ case, the Fast-DSCF decoder has significantly less area, and therefore is up to $5.8\times$ more area-efficient. In return, due to the iterative nature of the SCF algorithm, proposed decoder has larger worst-case latency than its SCL-based counterparts. 

As a result of the increased latency and power at higher error orders, it can be claimed that improving the FER costs more energy per decoded information bit; an increase of $37\%$ for $\omega=2$ and another $69\%$ for $\omega=3$ is observed. Since the synthesis results for the operating voltage and power consumption are not provided for the architectures described in \cite{FTSCF_ICC20,fastSSCL-TSP,Xia_SCL_TSP,Fan_SCL_JSAC}, we are not able to directly compare our results on power density and energy efficiency against the state-of-the-art. The only available results on energy consumption for SCL-based decooders have been carried out in \cite{ercan-allerton} for polar codes of shorter lengths (\textit{i.e.} $N=256$ and $N=512$), and it was shown that the increase in energy consumption is supralinear. For example, for the Fast-SSCL decoder for $PC(512,256)$, the difference in energy consumption between $L=2$ and $L=8$ is $6.68\times$. Note that, the energy consumption of Fast-SSCL decoding is the best among all the considered SCL-based algorithms in \cite{ercan-allerton}. Assuming that a similar increasing trend should be observed for the polar codes used in this work, the proposed Fast-DSCF decoder is expected to be much more energy-efficient than its fast SCL-based counterparts given the increase in energy consumption remains at $37\%$ between $\omega=1$ and $\omega=2$. 

The proposed Fast-DSCF implementations at $\omega>1$ have greater worst-case latency compared to their SCL-based counterparts. The worst-case latency increases with $T_{\text{max}}$, which is a major drawback for communications that prioritize low latency, including applications with FPGA. On the other hand, proposed implementations demonstrate improved area and energy efficiency as a trade-off, and have similar average latency and throughput compared to the state-of-the-art. Therefore, the proposed Fast-DSCF implementations are in favor of use cases where area and energy efficiency are prioritized over worst-case latency, such as massive machine-type communications \cite{5g_mmtc_IEEEComm2016}.

\section{Conclusion}\label{sec:conclusion}
In this work, we showed how to make the Dynamic SC-Flip decoding practical. We replaced the transcandental computations of the DSCF algorithm with a constant, and showed how to implement fast decoding techniques. Then, we further reduced the computational effort by employing a theoretical framework, which is later exploited in a hardware implementation. We showed further simplifications on metric updates and sorter complexity in hardware. Proposed approximations and simplifications do not alter the error-correction performance significantly but make the DSCF decoding practically feasible. The proposed Fast-DSCF decoders synthesized using TSMC $65$nm CMOS technology demonstrate $1.25$, $1.06$ and $0.93$ Gbps throughput for $\omega \in \{1,2,3\}$, respectively. For $\omega=3$, the Fast-DSCF decoder is up to $5.8\times$ more area-efficient than state-of-the-art fast CA-SCL decoders with equivalent FER performance at $L=16$. Compared to the state-of-the-art fast CA-SCL decoders with equivalent FER performance, the proposed decoders are up to $5.8\times$ more area-efficient. Finally, observations at energy dissipation indicate that the Fast-DSCF is more energy-efficient than its CA-SCL-based counterparts.

\section*{Acknowledgments}
The authors would like to acknowledge Prof. Dr. Jan Bajcsy from McGill University for his valuable insights on the derivation of the theoretical error probability on parity check codes.


\end{document}